\documentclass{aa}

\usepackage{graphicx}
\usepackage{txfonts}
\usepackage{hyperref}
\hypersetup{colorlinks, citecolor=blue}
\usepackage{soul, xcolor}
\setstcolor{red}

\begin{document}

\title{TDCOSMO}
\subtitle{XXII. Triaxiality and projection effects in time-delay cosmography}

\author{Xiang-Yu Huang
    \inst{1}, Simon Birrer\inst{1},
    Michele Cappellari\inst{2},
    Tommaso Treu\inst{3},
    Shawn Knabel\inst{3}
    \and
    Dominique Sluse\inst{4}
}

\institute{Department of Physics and Astronomy, Stony Brook University, Stony Brook, NY 11794, USA\\
    \email{xiangyu.huang.1@stonybrook.edu}\\
    \and
    {Sub-Department of Astrophysics, Department of Physics, University of Oxford, Denys Wilkinson Building, Keble Road, Oxford, OX1 3RH, UK}\\
    \and
    {Department of Physics and Astronomy, University of California, Los Angeles, CA 90095, USA}\\
    \and
    {STAR Institute, University of Li{\`e}ge, Quartier Agora, All\'ee du six Ao\^ut 19c, 4000 Li\`ege, Belgium}
}

\titlerunning{Triaxiality and projection effects in time-delay
    cosmography}
\authorrunning{Huang, XY. et al.}

\date{Received XXX; accepted YYY}

\abstract
{Constraining the mass-sheet degeneracy (MSD) is crucial for improving the precision and accuracy of time-delay cosmography.
    Joint analyses based on lensing and stellar kinematics have been widely adopted to break the MSD.
    A three-dimensional (3D) mass and stellar tracer population is required to accurately interpret the kinematics data.}
{Our forward-modeling procedure is aimed at evaluating the projection effects using strong lensing and kinematics observables and to determine an optimal model assumption for the stellar kinematics analysis leading to an unbiased interpretation of the MSD and $H_0$.}
{We numerically simulated the projection and selection effects for both a triaxial early-type galaxy (ETG) sample from the TNG100 simulation and an axisymmetric sample that matches the properties of slow-rotator galaxies representative of the strong lens galaxy population.
    Using the axisymmetric sample, we generated mock kinematics observables with spherically aligned axisymmetric Jeans anisotropic modeling (JAM) and assessed the kinematic recovery under different model assumptions.
    Using the triaxial sample, we quantified the random uncertainty introduced by modeling triaxial galaxies with axisymmetric JAM.
}
{
    We show that spherical JAM analysis of spatially unresolved kinematic data introduces a bias of up to 2\%-4\% (depending on the intrinsic shape of the lens) in the inferred MSD.
    Our model largely corrects this bias, resulting in a residual random uncertainty in the range of 0-2.2\% in the stellar velocity dispersion (0-4.4\% in $H_0$), depending on the  projected ellipticity and the anisotropy of the stellar orbits.
    This residual uncertainty can be further mitigated by the use of spatially resolved kinematic data, which constrain the intrinsic axis ratio. We also show that the random uncertainty in the kinematics recovery using axisymmetric JAM for axisymmetric galaxies is at the level of 0.24\% in the velocity dispersion, and the uncertainty using axisymmetric JAM for triaxial galaxies is at the level of 0.17\% in the velocity dispersion.
}
{}


\maketitle

%

\section{Introduction}\label{sec:intro}

Strong gravitational lenses are a powerful tool for probing absolute distances in the Universe and constraining its expansion history.
Time-delay measurements, which use the relative time delays between the multiple images of a time-varying source \citep{1964Refsdal}, provide precise constraints on the Hubble constant, $H_0$ \citep[e.g.,][hereafter time-delay cosmography, TDC]{Kundi_1997, Schechter_1997, 2003Koopmans, 2006Kochanek, 2006Saha, 2007Oguri, 2010Suyu, 2013Suyu}. Detailed reviews of TDC are presented in e.g., \citet{2016Treu_TDC, 2024Birrer}.

A key limiting factor on the constraining power of TDC is the mass sheet degeneracy \citep[MSD;][]{1985Falco}, which leaves the relative lensing observables invariant except for the time-delay prediction.
To acquire a precise and accurate measurement of the Hubble constant, the MSD must first be broken.
Currently, the primary method to break the MSD is to use stellar kinematics of the deflector galaxy to constrain the 3D gravitational potential of the deflector \citep{2002Treu&Koopmans, 2010Suyu, 2011Barnabe}.

For example, \citet{2020Birrer_TD4} (hereafter, TD4), exclusively constrained the MSD component of the deflector's density profile with measurements of the stellar velocity dispersion.
The measured Hubble constant by TD4 of the seven TDCOSMO lensed quasars is $H_0 = 74.5 ^{+5.6}_{-6.1}$ km s$^{-1}$ Mpc$^{-1}$, consistent with the $H_0 = 73.3 ^{+1.7}_{-1.8}$ km s$^{-1}$ Mpc$^{-1}$ value measured by the H0LiCOW collaboration \citep{2020Wong_H0licow}, assuming assertive mass profiles.
Since TD4 used a maximally flexible parameterization for the mass model with regard to the MSD in a Bayesian hierarchical framework, the uncertainty of the Hubble constant increases from two percent to eight percent, with the major component in the error budget being the precision in the kinematic observations and related modeling uncertainties.
To further improve the precision, TD4 imposed population prior from the stellar velocity dispersion profiles of 33 Sloan Lens ARCS (SLACS) lenses \citep{2006Bolton, 2015shu, 2021Shajib} and improved the uncertainty of $H_0$ to five percent , giving $H_0 = 67.4 ^{+4.1}_{-3.2}$ km s$^{-1}$ Mpc$^{-1}$; however, this value should be considered illustrative of the uncertainty rather than of the best estimate, owing to insufficient quality of the SDSS stellar velocity dispersions \citep{2025KM_kin_method}.
The key assumption of the TDCOSMO+SLACS analysis is that the external lens sample (i.e., the SLACS lenses) are from the same parent population as the TDCOSMO lenses.
Selection effects in the specific lens sample and across different samples must be studied and mitigated appropriately to the level of the stated precision in $H_0$ \citep[e.g.,][]{2024Sonnenfeld}.

Another assumption in TD4 is similar to that of previous studies, namely, that the kinematics interpretation assumed a spherical dynamical model, while real galaxies are in general nonspherical.
Therefore, the intrinsic shape and the orientation of the lens galaxy can be one potential source of systematic uncertainties, since the lensing and kinematics observables change with the viewing angle.
In this paper, we use the term "projection effect" to refer to the phenomenon that the lensing and kinematics observables of a galaxy depend on the orientation relative to the observer.
For example, an oblate lens galaxy which appears edge-on to an observer will have more projected mass-density at its center and therefore will have a larger Einstein radius, while with a face-on orientation, it will have the smallest Einstein radius.
For a spherical galaxy of the same total mass, the Einstein radius falls between the maximum and minimum Einstein radius created by the oblate galaxy.
For lens mass modeling, the deprojection of the lens galaxy is unnecessary since all lensing quantities are determined by the projected mass-density.
In contrast, kinematics modeling, which is used to constrain the 3D mass distribution and break the MSD, requires the deprojection of the lens galaxy in light and mass \citep[e.g., ][]{2011Dutton}.

For an individual galaxy, the deprojection for the lens mass from photometric data is in general underconstrained.
The deprojection for the kinematics model is possible through integral field spectroscopy \citep[IFS, e.g., ][]{2008Cappellari} only under certain assumptions on the stellar kinematics anisotropy.
Population-level deprojection of galaxies are more applicable, since the intrinsic shape distribution can be obtained by statistically inverting the distribution of projected ellipticities of a sample of galaxies \citep{1970Sandage, 1980Benacchio, 1981Binney, 1992Lambas, 1992Ryden, 2005Vincent, 2007Kimm, 2008Padilla, 2013Rodriguez} under the assumption of isotropy of the sample; alternatively, another approach is to invert the distribution of the projected ellipticities and the misalignment angle between the photometric and the kinematic axes \citep{2014Weijmans, 2018Li, 2018Ene}, despite obtaining potentially nonunique solutions.
The intrinsic shape distribution of galaxies can be used to model the projection effects and selection effects under some selection criterion.
To do so, a thorough understanding of the intrinsic galaxy population acting as strong lenses is required.

Most strong gravitational lenses discovered so far are massive early-type galaxies (ETGs), which are the most massive galaxies in the Universe.
This is a direct consequence of the lensing cross-section.
The images of the background source must have large enough angular separations to be identified as multiply lensed images. In regard to their dynamical structures,
\cite{2011Barnabe} showed that a sample of early-type galaxies from the SLACS Survey can be divided into the two usual kinematics classifications, slow rotators and fast rotators \citep[e.g., ][]{Emsellem2007,Cappellari2016}. 
More specifically, \citet{2018Li} modeled the projected shape of a complete sample of all 189 massive ($M_*>2\times10^{11} M_\odot$) slow rotator ETGs out of a sample of 2200 galaxies from the DR14 of the SDSS MaNGA IFU survey.
They found a weakly triaxial shape consistent with a dominant triaxial-oblate population under a wide set of model assumptions and no significant fraction of prolate-triaxial galaxies.
Fast-rotator ETGs, on the other hand, are found to be much flatter and oblate-like \citep{2014Weijmans}, showing consistency with their dynamical status.
For either type of ETG, kinematics modeling with state-of-the-art Jeans anisotropic modeling (JAM) method \citep{2008Cappellari, 2020Cappellari} can aptly recover the 3D kinematics model, which is  consistent (in projection) with the observational data, under certain assumptions on the intrinsic shape of the ETG (e.g., assuming spherical or axial symmetry of the gravitational potential).
In the axisymmetric case, the projection effect of the 3D kinematics model will impact the interpretation of the kinematics in TDC.
The projection effect in the kinematics model also leads to potential selection bias under lens selection criterion in the lens finding stage and follow-up analysis, due to the change in lensing efficiency with orientation.

The goal of this paper is to investigate the projection and selection effects introduced by the intrinsic shape of the mass distribution and the kinematics stellar tracer distribution of the lensing galaxies. As the projection effect affects the observation and interpretation of lensing and kinematics observables, we aim to assess qualitatively and quantitatively its impact on the measurement of the Hubble constant via TDC. For context, we note that uncertainty or bias in the stellar velocity dispersion $\sigma$ affects the determination of the Hubble $H_0$ as $\delta H_0/H_0=2\delta \sigma / \sigma$. Therefore, if we aim to measure $H_0$ with a 2\% accuracy, biases on $\sigma$ predicted by the lens model need to be contained below 1\%.

To achieve our goal, we study the projection and selection effects in the lensing observables using two galaxy samples: one is a synthetic axisymmetric sample, which resembles the SLACS lenses, and the other is a triaxial ETG sample from the IllsutrisTNG-100 simulation.
We created a catalog of mock lensing observables (e.g., projected ellipticities and the Einstein radii) for both samples, assuming the viewing angle is random.
We quantified the projection effect of individual galaxies with the scatters in the Einstein radius.
We qualitatively analyzed the selection effect by applying a selection criterion on the Einstein radius.
On the kinematics side, using the synthetic axisymmetric sample, we created mock kinematics observables and analyzed the projection and selection effects specific to axisymmetric systems.
We then quantified the accuracy in the kinematics recovery under different assumptions on the galaxy model with JAM.
Specifically, we explicitly derived the quantitative corrections in the interpretation of the kinematics observables as a function of measured projected ellipticity and estimated the residual uncertainty for the corrected velocity dispersion. We show that it is possible to constrain potential $H_0$ biases below the target 1\% precision level.
We review the key axisymmetric assumption in the kinematics modeling by comparing the kinematics observables from the axisymmetric kinematics models of triaxial systems.  We also obtain a numerical calculation of an upper limit of the relative difference in the velocity dispersion using the triaxial TNG ETG sample, which is more elliptical and triaxial than real lensing galaxies such as the SLACS lenses at the population level.

This paper is organized as follows. Section~\ref{sec:theory} discusses the theoretical basis including lensing basics, projection formalism, and kinematics modeling.
Section~\ref{sec:sample_construction} discusses our lens sample construction.
Section~\ref{sec:proj_effect_lensing} shows the projection and selection effect in the lensing observables of a triaxial catalog of ETGs from the TNG100 simulation.
In Section~\ref{sec:kinematics_results}, we  describe how we generated mock kinematics data using axisymmetric JAM for a synthetic axisymmetric lens sample presented in Section \ref{sec:sample_construction}. Then, we discuss the selection effect for axisymmetric systems.
We also compare the kinematics model assumptions by estimating their relative biases in the kinematics recovery.
In Section \ref{sec:triaxiality_kinematics}, we discuss the effect of triaxiality in the kinematics recovery of ETG using axisymmetric JAM.
We present our conclusions in Section~\ref{sec:conclusion}. Throughout this work, we assume a flat $\Lambda$CDM cosmology with $H_0 = 70$ km s$^{-1}$ Mpc$^{-1}$, where required.
We note that this arbitrary choice does not affect any of our conclusions regarding the selection function and quantitative correction terms.

\section{Basics of time-delay cosmography}\label{sec:theory}

This section reviews the theoretical basis of this work, including strong lensing observables and formalism (Section~\ref{subsec:lensing_formalism}), the mass sheet degeneracy (Section~\ref{subsec:intro_msd}), projection of triaxial systems (Section~\ref{subsec:projection_formalism}), kinematics modeling with JAM (Section~\ref{subsec:kinematics_formalism}), and the projection formalism of axisymmetric multi-Gaussians (Section~\ref{subsec:projection_form_axisymmetric}).

\subsection{Strong lensing formalism}\label{subsec:lensing_formalism}

The deflection of light follows the lensing equation,
\begin{equation}\label{eqn:lens_equation}
    \boldsymbol{\beta} = \boldsymbol{\theta} - \boldsymbol{\alpha}(\boldsymbol{\theta}),
\end{equation}
where $\boldsymbol{\beta}$ is the unlensed angular source position, $\boldsymbol{\theta}$ is the angular image position seen from the observer, and $\boldsymbol{\alpha}$ is the deflection angle as seen on the sky.
The deflection angle can be expressed as the gradient of the lensing potential
\begin{equation}
    \boldsymbol{\alpha}(\boldsymbol{\theta}) = \nabla\psi(\boldsymbol{\theta}),
\end{equation}
which is related to the total projected mass-density on the lens plane via
\begin{equation}
    \kappa(\boldsymbol{\theta}) = \frac{1}{2} \nabla^2 \psi.
\end{equation}
Here $\kappa(\boldsymbol\theta)$ is the convergence, defined as
\begin{equation}\label{eqn:def_kappa}
    \kappa (\boldsymbol\theta) = \frac{\Sigma(\boldsymbol\theta)}{\Sigma_{\mathrm{crit}}}
\end{equation}
under small angle approximation.
$\Sigma(\boldsymbol\theta)$ is the total projected excess mass-density compared to the cosmological background density on the lens plane, and $\Sigma_\mathrm{crit}$ is the critical density for the lens-source configuration
\begin{equation}
    \Sigma_{\mathrm{crit}} = \frac{c^2}{4\pi G}\frac{D_\mathrm{s}}{D_\mathrm{ds}D_\mathrm{d}}.
\end{equation}
Here, $D_\mathrm{s}$, $D_\mathrm{d}$, and $D_\mathrm{ds}$ are the angular diameter distances to the source, to the deflector, and from the deflector to the source. The strong lensing efficiency, characterized by the deflection angle $\boldsymbol\alpha$, is determined by the total projected mass within the Einstein radius $\theta_E$.
The Einstein radius, $\theta_E$, is the angular radius within which the mean convergence is unity:
\begin{equation}\label{eqn:def_theta_e}
    \int_{\mathcal{A}(<\theta_{\rm E})} \kappa(\boldsymbol\theta) d\boldsymbol\theta\equiv \mathcal{A}(<\theta_{\rm E}).
\end{equation}
Here, $\mathcal{A}(<\theta_{\rm E})$ is an area with $\theta_E$ as the effective radius $\mathcal{A}(<\theta_{\rm E}) \equiv \pi \theta_{\rm E}^2$.
We specify here that the definition of the Einstein radius in this paper is the circularized value over the lens plane, i.e., $\sqrt{q'}\theta_E^\mathrm{maj}$, where $q'$ is the apparent axis ratio of the iso-density contours, and $\theta_E^\mathrm{maj}$  is the Einstein radius measured along the major axis of the elliptical iso-density contours.

The time delays between the different images are proportional to the difference in their Fermat potential:
\begin{equation}
    \Delta t_{AB} = \frac{1}{c}D_{\Delta t}\left [\phi (\boldsymbol{\theta}_A, \boldsymbol\beta) - \phi (\boldsymbol{\theta}_B,\boldsymbol\beta)\right],
\end{equation}
where $c$ is the speed of light, A and B label two different images of the same background source, $\phi (\boldsymbol\theta, \boldsymbol\beta)$ is the Fermat potential
\begin{equation}
    \phi (\boldsymbol\theta, \boldsymbol\beta) = \frac{(\boldsymbol\theta - \boldsymbol\beta)^2}{2} - \psi(\boldsymbol\theta),
\end{equation}
and $D_{\Delta t}$ is a coefficient with the dimension of distance, also called the time-delay distance
\begin{equation}
    D_{\Delta t} = (1 + z_\mathrm{d}) \frac{D_\mathrm{d} D_{\mathrm{s}}}{D_{\mathrm{ds}}}.
\end{equation}
Then, with a measurement of the time delay $\Delta t$ and an inference on the difference of the Fermat potential, one can measure the time-delay distance.
The Hubble constant $H_0$ scales inversely with the angular diameter distance, and thus
\begin{equation}
    H_0 \propto D_{\Delta t}^{-1}.
\end{equation}

\subsection{The mass sheet degeneracy}\label{subsec:intro_msd}
The mass sheet degeneracy \citep[MSD;][]{1985Falco} stems from the so-called mass sheet transform (MST) on the convergence and the unknown source position
\begin{align}
    \kappa(\boldsymbol\theta) & \rightarrow \kappa_{\lambda}(\boldsymbol\theta) = \lambda\kappa(\boldsymbol\theta) + 1 - \lambda, \label{eqn:def_msd1} \\
    \boldsymbol\beta          & \rightarrow \boldsymbol\beta_{\lambda} = \lambda \boldsymbol\beta.\label{eqn:def_msd2}
\end{align}
Here, $\boldsymbol \theta$ is the angular position of the lensed image of the source and $\boldsymbol\beta$ is the angular position of the unlensed source, as in Eq. (\ref{eqn:lens_equation}).
Here $(1 - \lambda)$ acts as an infinite sheet of mass. The MST preserves the image configuration and the relative magnification between the multiple images of the source, but changes the time delay prediction under a fixed lens mass model by
\begin{equation}
    \Delta t_\lambda = \lambda \Delta t.
\end{equation}
The time-delay distance is then transformed via
\begin{equation}
    D_{\Delta t, \lambda} = \lambda^{-1} D_{\Delta t}.
\end{equation}
The inferred Hubble constant then changes via
\begin{equation}
    H_{0, \lambda} = \lambda H_0.
\end{equation}

The MSD effect can be divided into two components regarding the source of the over- or under-density with respect to the background: (a) the internal MSD $\lambda_\mathrm{int}$, which is associated with the mass distribution of the deflector galaxy due to changes in the radial density profile; and (b) the external MSD, contributed by the projected mass of the line-of-sight (LoS) structure,  $\kappa_\mathrm{ext}$, outside the associated halo.

The total MSD  \citep{2013Schneider&sluse, 2020Birrer_TD4}
can be expressed as\begin{equation}
    \lambda \approx (1 - \kappa_\mathrm{ext})\lambda_\mathrm{int}.
\end{equation}
Both the internal and the external MSD have an effect on the deflector potential.
To first order, the total MSD can be constrained with stellar kinematics.
In this work, we mainly focus on the projection effect caused by the surface density of deflector galaxy itself; thus, we refer to $\lambda \approx \lambda_\mathrm{int}$ for the remainder of this paper, noting that the external component has an additional impact on the kinematics \citep[e.g.,][]{2006Fassnacht}.

\subsection{Projection effect and projection formalism for triaxial systems}\label{subsec:projection_formalism}

According to Eq. (\ref{eqn:def_kappa}), all lensing quantities, including the Einstein radius, can be derived from the 2D surface mass-density,
\begin{equation}\label{eqn:trivial_int_los}
    \Sigma\left(x', y'\right) =\int_{-\infty}^{\infty} \rho\left(x', y', z'\right) d z'.
\end{equation}
Here $(x', y')$ are coordinates on the lens plane and $z'$ is the third dimension along the LoS.
If the intrinsic mass-density distribution is nonspherical, $\Sigma(x', y')$ is dependent on the LoS direction.
In this paper, the term "projection effect" specifically refers to the change in the surface mass-density, light surface brightness, and the stellar kinematics under different viewing angles of the lens galaxy.

The projection formalism for density profiles stratified on similar concentric spheroids is presented in detail by \cite{1977Stark} and \cite{1985Binney}.
Here we follow the convention of \cite{1985Binney} and briefly summarize it.
We assume the mass-density profile of a triaxial galaxy has a constant radial shape, without angular twists or changes in the axis ratio as a function of radius. With this assumption, the density is only a function of the ellipsoidal radius variable, namely,
\begin{equation}
    \rho = \rho(a_v),
\end{equation}
with
\begin{equation}\label{eqn:norm_factor_introduction}
    a_v = Z \sqrt{x^2 + \frac{y^2}{p^2} + \frac{z^2}{q^2}}.
\end{equation}
Here, the ellipsoid has its major, intermediate, and minor axis aligned with the $x$, $y,$ and $z$ axis of the Cartesian coordinate system, respectively;
$p$ is the axis ratio between the intermediate and major axis of the ellipsoid and $q$ is the axis ratio between the minor and major axis of the ellipsoid.
$1 \geq p \geq q$.

The coefficient $Z$ is a normalization factor added to preserve mass when varying the shape parameters $p$ and $q$.
There are two choices to preserve the total integrated mass when varying the shape parameters: (a) $Z=1$ (not rescaling the radius) and renormalize the density uniformly in $\rho(a_v)$, and (b) $Z= (pq)^{1/3}$ without any renormalization of the overall density.
A detailed comparison of the two choices for the density normalization is discussed in Appendix \ref{apdx:norm_factor}.

When a galaxy is viewed along a LoS characterized by the polar angle $\theta$ and the azimuthal angle $\phi$ in a spherical coordinate system, its projected mass-density in Eq. (\ref{eqn:trivial_int_los}) can be written as
\begin{equation}
    \Sigma\left(x', y'\right) =\frac{2}{\sqrt{f}} \int_0^{\infty} \rho\left({z''}^{2}+a_{s}^2\right) d z''.
\end{equation}
Here, $z''$ is an integration variable, and $a_s$ is the elliptical radius variable of the iso-density contour in projection,\begin{equation}
    a_{s}^2=\frac{1}{f}\left[A {x'}^{2}+B x' y'+C {y'}^{2}\right].
\end{equation}Parameters $A$, $B$, $C,$ and $f$ are solely determined by the LoS direction ($\theta$, $\phi$) and the intrinsic axis ratios $p$ and $q$ as expressed in Eqs. (6, 11) in \cite{1985Binney}, namely,
\begin{align}
    A = & Z^2\left(\frac{\cos ^2 \theta}{q^2}\left(\sin ^2 \phi+\frac{\cos ^2 \phi}{p^2}\right)+\frac{\sin ^2 \theta}{p^2}\right), \\
    B = & Z^2\cos \theta \sin 2 \phi\left(1-\frac{1}{p^2}\right) \frac{1}{q^2},                                                    \\
    C = & Z^2\left(\frac{\sin ^2 \phi}{p^2}+\cos ^2 \phi\right) \frac{1}{q^2},                                                     \\
    f = & Z \left(\sin ^2 \theta\left(\cos ^2 \phi+\frac{\sin ^2 \phi}{p^2}\right)+\frac{\cos ^2 \theta}{q^2}\right).
\end{align}

The apparent axis ratio $q' (q' \leq 1$) of the projected elliptical iso-density contour is
\begin{equation}\label{eqn:def_projcted_q}
    q'(\theta, \phi , p, q)=\sqrt{\frac{A+C-\sqrt{ (A-C)^2+B^2}}{A+C+\sqrt{ (A-C)^2+B^2}}}.
\end{equation}
We define the projected ellipticity of the surface mass-density as $e(\theta, \phi , p, q) = (1 - q')/(1 + q')$.

The projection formalism also works for the stellar luminosity component of the galaxy, trading the intrinsic mass-density, $\rho(a_v),$ for the intrinsic luminosity distribution, $l(a_v)$.

\subsection{Stellar kinematics modeling}\label{subsec:kinematics_formalism}

Stellar kinematics data provide an independent measurement of the gravitational potential of the deflector galaxy and thus can break the MSD.
The luminosity-weighted LoS velocity dispersion  $\sigma^P$ is effective in constraining the mass-density slope of elliptical galaxies via joint lensing and kinematics analysis under certain assumptions on the stellar anisotropy profile \citep{2002Treu&Koopmans}.
For a given stellar tracer distribution, the squared velocity dispersion is proportional to the mass-density or, equivalently, the mass-to-light ratio, $M/L$.
For $\lambda\approx1$ and near the galaxy center $\boldsymbol\theta\approx0$ where the surface density is large, the first term in Eq. (\ref{eqn:def_msd1}) dominates over the second and thus $\kappa_{\lambda}(\boldsymbol\theta) \approx \lambda\kappa(\boldsymbol\theta)$.
In this limit, the MST has the same effect as varying the total $M/L$, and the velocity dispersion is transformed via
\begin{equation}\label{eqn:sigma_under_mst}
    \sigma^P_\lambda \approx \sqrt{\lambda} \sigma^P.
\end{equation}
This is equivalently $\sigma^P_* \approx \sqrt{\lambda} \sigma^P_l$, where $\sigma^P_*$ is the stellar velocity dispersion and $\sigma^P_l$ is the modeled velocity dispersion derived from the best fit lens model of the imaging data.
The inferred $\lambda$ from comparing observed and predicted velocity dispersion can then be applied to correct the lensing potential and, hence, $H_0$ in the inference \citep{2020Birrer_TD4}.
The relative uncertainty in constraining $H_0$ that comes from the uncertainty in the velocity dispersion (model or measurement) is transformed via
\begin{equation}
    \frac{\delta H_0}{H_0} = \frac{\delta \lambda}{\lambda}=2 \frac{\delta \sigma^{\mathrm{P}}}{\sigma^{\mathrm{P}}}.
\end{equation}

The kinematic modeling methods for galaxies can be grouped into three categories. First, we have particle-based methods \citep[e.g., made-to-measure;][]{1996Syer} and orbit-based methods \citep[e.g., Schwarzchild;][]{1979Schwarzschild} are highly flexible and can describe complex systems, including triaxial and time-dependent configurations. They fit the data with sets of orbital or particle weights. These approaches are particularly powerful when high-quality kinematic data provide higher-order Gauss-Hermite moments beyond the mean velocity and velocity dispersion. The main drawbacks are computational cost, especially for joint lensing and kinematics inference, and discreteness noise. Most importantly, the stellar distribution function (DF) is represented implicitly by the fitted weights. This implicit form makes the methods unsuitable for tasks that require an explicit, parameterized DF, such as constructing mock datasets. Second, we have analytic DF-based solutions \citep[e.g., ][]{2005Mamon} provide explicit functional forms. They are well suited to direct analysis and forward modeling, but their applicability is usually limited to simple symmetries, such as spherical or basic axisymmetric configurations. Third, we have solutions of the Jeans equations (Jeans 1922), in particular, the Jeans anisotropic modeling (JAM) method \citep{2008Cappellari,2020Cappellari}, which offer an effective alternative. JAM solves the Jeans equations for axisymmetric or spherical systems and allows for velocity anisotropy. It uses multi-Gaussian expansion \citep[MGE; ][]{1994Emsellem, 2002Cappellari}(MGE; Emsellem et al. 1994; Cappellari 2002) to approximate the mass and tracer densities. Because JAM provides an explicit, parameterized description of the kinematic moments and the anisotropy, it enables direct computations of the observables and the efficient construction of mock datasets, as required for the present work.

In this work, we use the spherically aligned axisymmetric JAM framework \citep{2020Cappellari}, which has the velocity ellipsoid aligned with a spherical coordinate system, for the dynamical modeling of stellar kinematics observables. Compared with the cylindrically aligned JAM proposed in \citet{2008Cappellari}, the spherical-aligned JAM provides a better estimate of the stellar kinematics for slow-rotator type ETGs, as most of the strong gravitational lenses are of this type \citep{2024Knabel}, because the slow-rotator ETG population are found to have a generally rounder or weakly triaxial shape \citep{Cappellari2016, 2018Li}.

We present a brief description of the formalism of spherically aligned JAM.
The stellar motion in a galaxy under equilibrium can be described with a six-dimensional (6D) distribution function (DF) $f(\boldsymbol x, \boldsymbol v)$, which follows the steady-state, collisionless Boltzmann equation under a smooth gravitational potential $\Phi(\boldsymbol x)$ \citep[Eq.~4-13b, ][]{1987BT}, expressed as
\begin{equation}\label{eqn:boltzmann_eqn}
    \sum_{i=1}^3\left(v_i \frac{\partial f}{\partial x_i}-\frac{\partial \Phi}{\partial x_i} \frac{\partial f}{\partial v_i}\right)=0.
\end{equation}
Equation (\ref{eqn:boltzmann_eqn}) can be solved under some simplifying assumptions: here we assume axial symmetry, namely, $\partial \Phi/\partial \phi =0$, $\partial f/\partial \phi =0$, where $\phi$ is the polar angle in a spherical coordinate, with the azimuthal angle, $\theta =0,$ aligned with the symmetry axis. We also assume that the velocity ellipsoid is aligned with the spherical coordinate, i.e., $\langle v_r v_\theta \rangle=0$.
Eq.~(\ref{eqn:boltzmann_eqn}) is then reduced to \citep{1996dezeeuw, 2020Cappellari}
\begin{align}
    \frac{\partial\left(\rho_* \langle v_r^2 \rangle\right)}{\partial r}+\frac{(1+\beta) \rho_* \langle v_r^2 \rangle-\rho_* \langle v_\phi^2\rangle}{r}=-\rho_* \frac{\partial \Phi}{\partial r}, \label{eqn:axi_jeans_1} \\
    (1-\beta) \frac{\partial\left(\rho_* \langle v_r^2\rangle \right)}{\partial \theta}+\frac{(1-\beta) \rho_* \langle v_r^2\rangle -\rho_* \langle v_\phi^2\rangle}{\tan \theta}=-\rho_* \frac{\partial \Phi}{\partial \theta}\label{eqn:axi_jeans_2},
\end{align}
which are the so-called Jeans \citep{1922Jeans} equations under axial symmetry.
Here $\rho_*$ is the density of the dynamical tracer,
\begin{equation}
    \rho_* = \int f(\boldsymbol x, \boldsymbol v) d\boldsymbol v,
\end{equation}
$\rho_* \langle v_i v_j\rangle$ are the elements of the stress tensor,
\begin{equation}
    \rho_* \langle v_i v_j\rangle = \int v_i v_j f(\boldsymbol x, \boldsymbol v) d\boldsymbol v,
\end{equation}
and $\beta$ is defined as the anisotropy parameter in the spherical coordinate,
\begin{equation}
    \beta = 1 - \frac{\langle v_\theta^2\rangle}{\langle v_r^2\rangle}.
\end{equation}
Using a set of multi-Gaussian decomposition \citep{1994Emsellem, 2002Cappellari} for the mass-density generating the gravitational potential and the intrinsic density of the dynamical tracer, and by assuming a spatially constant anisotropy parameter $\beta$ and a constant mass-to-light ratio, $M/L$, for each Gaussian, JAM solves Eq. (\ref{eqn:axi_jeans_1}-\ref{eqn:axi_jeans_2}) and predicts the LoS luminosity-weighted second moment of the velocity
\begin{equation}
    \Sigma\langle (\sigma^P)^2 \rangle (x', y') = \int \rho_* \langle v_z'^2\rangle dz',
\end{equation}
where $\Sigma (x', y')$ is the surface brightness of the galaxy.

The observed second moment of velocity is affected by the atmospheric seeing and the instrumental response.
The prediction for the observed second moment within an aperture, $\mathcal{A}$, is then the aperture-integrated, luminosity-weighted, seeing-convolved dispersion \citep{2010Suyu}
\begin{equation}\label{eqn:def_sigma_p}
    \langle (\sigma^P)^2 \rangle_\mathcal{A} = \frac{\int_{\mathcal{A}}\left[\Sigma(x',y') \langle (\sigma^P)^2 \rangle * \mathcal{P}\right] \mathrm{d}x'\mathrm{d}y'}{\int_{\mathcal{A}}\left[\Sigma(x',y') * \mathcal{P}\right] \mathrm{d}x'\mathrm{d}y'},
\end{equation}
where $* \mathcal{P}$ denotes a convolution with the PSF $\mathcal{P}$.

\subsection{Axisymmetric MGE projection formalism}\label{subsec:projection_form_axisymmetric}

The projection formalism for axisymmetric systems, where the axisymmetric Jeans equations can be solved using the MGE parameterization of the stellar light profile, can be simplified from the projection formalism described in Sect. \ref{subsec:projection_formalism} by taking $p=1$ for oblate systems, and $p=q$ for prolate systems.
However, when using the MGE parameterization, the projected axis ratio follows a simpler calculation and the projected profile can be calculated using the conservation of total luminosity instead of the brute-force numerical integrals.
Here, we briefly summarize the projection formalism, following the derivation in \citet{2020Cappellari}.

The intrinsic luminosity profile of a galaxy $\nu(r, \psi)$ can be expressed with
\begin{equation}\label{eqn:density_mge_expression}
    \nu(r, \psi)=\sum_{k=1}^N \nu_{0 k} \exp \left[-\frac{r^2}{2 \sigma_k^2}\left(\sin ^2 \psi+\frac{\cos ^2 \psi}{q_k^2}\right)\right],
\end{equation}
where $\psi$ is the angle between the field point $(r, \psi)$ and the symmetry axis of the ellipsoid, $r$ is the distance to the galaxy center, $\nu_{0k}$,  $\sigma_k$ and $q_k$ are the peak, dispersion, and the intrinsic axis ratio of the $k$-th MGE component.
The intrinsic axis ratio $q_k$ is $q_k < 1$ for oblates and $q_k>1$ for prolates.
The surface brightness profile $\Sigma(x', y')$ is then
\begin{equation}
    \Sigma\left(x', y'\right)=\sum_{k=1}^N \Sigma_{0 k} \exp \left[-\frac{1}{2 \sigma_k^2}\left(x'^{2}+\frac{y'^{2}}{{q'}_k^{2}}\right)\right],
\end{equation}
where $x'$ is aligned with the photometric major axis.
Similar to the notation of $q_k$, $q'_k$ has $q'_k < 1$ for oblates, and $q'_k>1$ for prolates.

The orientation of an axisymmetric system is characterized by the inclination angle, $i$, which is the angle between the LoS and the symmetry axis of the axisymmetric ellipsoid.
For isotropic inclination angles, $\cos i \in \mathcal{U}[0, 1]$.
The the projected axis ratio, $q'$, for an oblate galaxy is computed directly with
\begin{equation}\label{eqn:axis_ratio_proj_oblate}
    {q'}^2_k = q^2_k \sin^2 i + \cos^2 i,
\end{equation}
where the subscript $k$ labels the k-th MGE component. For a prolate galaxy,
\begin{equation}\label{eqn:axis_ratio_proj_prolate}
    \frac{1}{{q'}^2_k} = \frac{\sin^2 i}{q^2_k} + \cos^2 i.
\end{equation}

The total mass or total luminosity of each MGE component is conserved in the projection, leading to
\begin{equation}\label{eqn:peak_density_proj}
    \Sigma_{0k} = \frac{q_k \nu_{0k} \sigma_k \sqrt{2\pi}}{q'_k}.
\end{equation}

\section{Lens sample construction}\label{sec:sample_construction}

In this section, we discuss our lens sample construction. 
We used two different lens sets for the analysis of the projection effect and the selection effect in lensing and kinematics. For the analysis on the kinematics, we used a synthetic axisymmetric lens sample similar to the SLACS lenses in the $R_e-\sigma_v$ space, with empirical intrinsic shape distributions. For the lensing analysis, we used a triaxial ETG sample from the TNG100 simulation. We also used this triaxial sample for the analysis of the triaxiality in axisymmetric kinematics models (Section~\ref{sec:triaxiality_kinematics}).
In Section \ref{subsec:lens_sample}, we describe these two samples.
In Section \ref{subsec:lens_gal_model}, we describe our choice of the mass, light, and anisotropy profiles for the lens samples.
In Section \ref{subsec:lens_gal_kin_model}, we list the simplifying assumptions on the kinematics models of our sample.

\subsection{Sample properties}\label{subsec:lens_sample}

\subsubsection{The synthetic lens set}

\begin{figure}
    \centering
    \includegraphics[width = \linewidth]{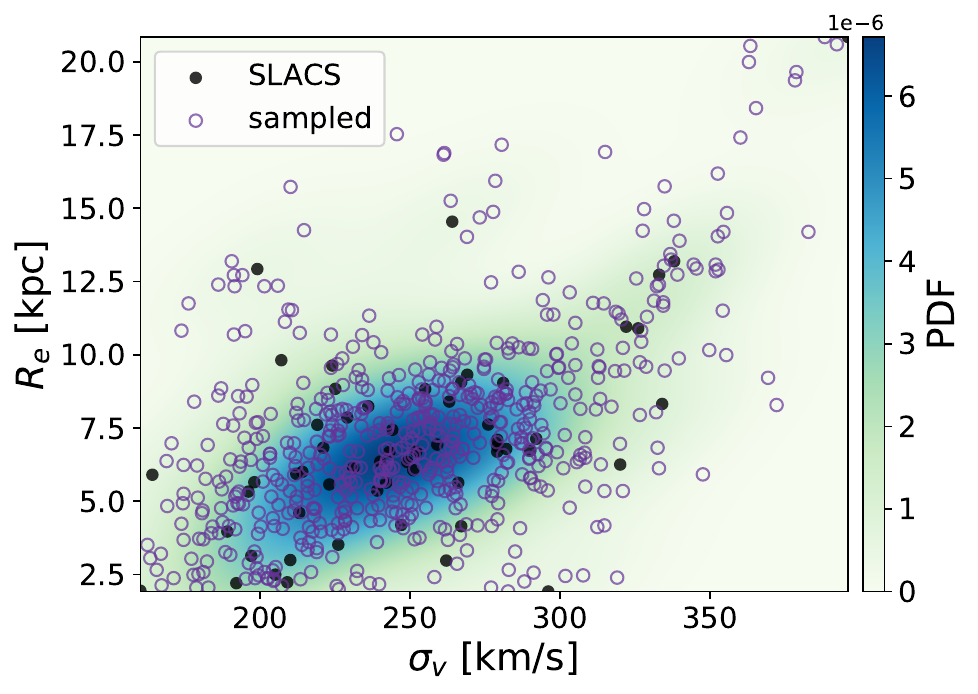}
    \caption{Axisymmetric synthetic lens set in the $\sigma_v - R_e$ space. Black dots are the SLACS grade "A" lenses reported in \citet{2008Bolton}. The purple circles are drawn from the 2D KDE of the original data points. }
    \label{fig:synthetic_slacs}
\end{figure}

We generated an axisymmetric synthetic lens sample using priors from the SLACS lenses \citep{2008Bolton}.
We used the velocity dispersion of the SLACS as a description of the total mass normalization, along with the effective radius as a description of the shape of the stellar tracer profile. In practice, we sample velocity dispersion and effective radius from the 2D kernel density estimator (KDE) constructed using the grade "A" lenses in Table 4 of \citet{2008Bolton}.
Figure \ref{fig:synthetic_slacs} shows the 2D KDE, the original data points of the SLACS lenses in the $\sigma_v - R_e$ space, and the synthetic lens sample.
The size of the synthetic lens sample was set as 600.

We then assigned to each lens an intrinsic axis ratio drawn from $\mathcal{N}[\mu=0.74, \sigma=0.08]$ for the oblate sample and $\mathcal{N}[\mu=0.84, \sigma=0.04]$ for the prolate sample.
These values are inferred from the SDSS MaNGA slow-rotator ETG \citep[Table 1 of ][line 1 for the oblate shape distribution, and line 4 for the prolate shape distribution]{2018Li} using the relation between the misalignment angles between the kinematics major axis and the photometric major axis, and the observed ellipticities.

\subsubsection{The TNG100 triaxial ETG}\label{subsubsec:tng_intro}

We also used a triaxial ETG catalog from the TNG100 simulation to illustrate the projection effect of triaxial systems.
The IllustrisTNG is a set of large-scale, gravity+magnetohydrodynamical simulations for the study of galaxy formation and evolution \citep{2018Pillepich}, including three primary runs with different volumes and resolutions, the TNG300, the TNG100, and the TNG50, with side length $\sim$ 300, 100, and 50 Mpc, respectively.
The catalog we used was also selected in \citet{pulsoni_stellar_2020},  spanning a stellar mass range of $10^{10.3}$ - $10^{12} $M$_{\odot}$, used to study the relation between the stellar kinematics and the intrinsic shape at different radial scales.
The selection was performed on snapshots at $z = 0$ in the color-stellar mass diagram, in which the selected ETGs are confirmed to match with the observed ETGs in IFU surveys such as the SDSS MaNGA \citep{2015Bundy} and SAMI \citep{2012Croom}.
For a detailed description of their selection, we refer to Section~5 of \cite{pulsoni_stellar_2020}. For the selected sample, the physical properties were extracted from the simulated photometric and IFU data.  We used the following in this work:
\begin{itemize}

    \item Effective radius in projection, $R_e$: half-mass radius of the total bound stellar mass on the projected major axis when the projection is along one random LoS.
          For the remainder of this paper, we assume that $R_e$ from the TNG catalog is the circularized half-light radius in projection, namely, $R_e^\mathrm{circ} = \sqrt{q'}R_{e}^{\mathrm{major}}$, where $q'$ is the projected apparent axis ratio.

    \item Velocity dispersion at effective radius, $\sigma_\mathrm{rm}$: velocity dispersion map is obtained for a random LoS. The mean velocity and velocity dispersion of stellar particles are calculated within a radial bin (annulus) around $R_e$. We specify here that we did not adopt $\sigma_\mathrm{rm}$ as the "true" velocity dispersion of the sample galaxies, but as a quantity to represent the total mass normalization depending on the mass-density profile of the lens galaxies.

    \item Intrinsic axis ratios: the intrinsic intermediate-to-major axis ratio, $p$,  and minor-to-major axis ratio $q$ characterizing the iso-density contours of each galaxy. It was calculated from the inertia tensor, $I_{ij} = \sum_n x_{n,i}x_{n,j}M_n / \sum_n M_n$, in each radial bin, where the summation $\sum_n$ was performed on the 50\% nearest stellar particles, $x_{n,i}$ are the coordinates, and $M_n$ are their mass.
          For simplicity of our strong lensing mock tests, we ignored the radial variation of $p$ and $q$ in the simulation and used the $p$ and $q$ value at $1R_e$ as constant axis ratios throughout all radii.
          We also assumed a perfect alignment between the stellar mass distribution,  stellar luminosity, and  total mass distribution; therefore, in the next steps of the study, $p$ and $q$ were used to characterize the shape of the stellar tracer as well as the total density.
\end{itemize}

\begin{figure}[htbp]
    \centering
    \includegraphics[width = 0.95\linewidth]{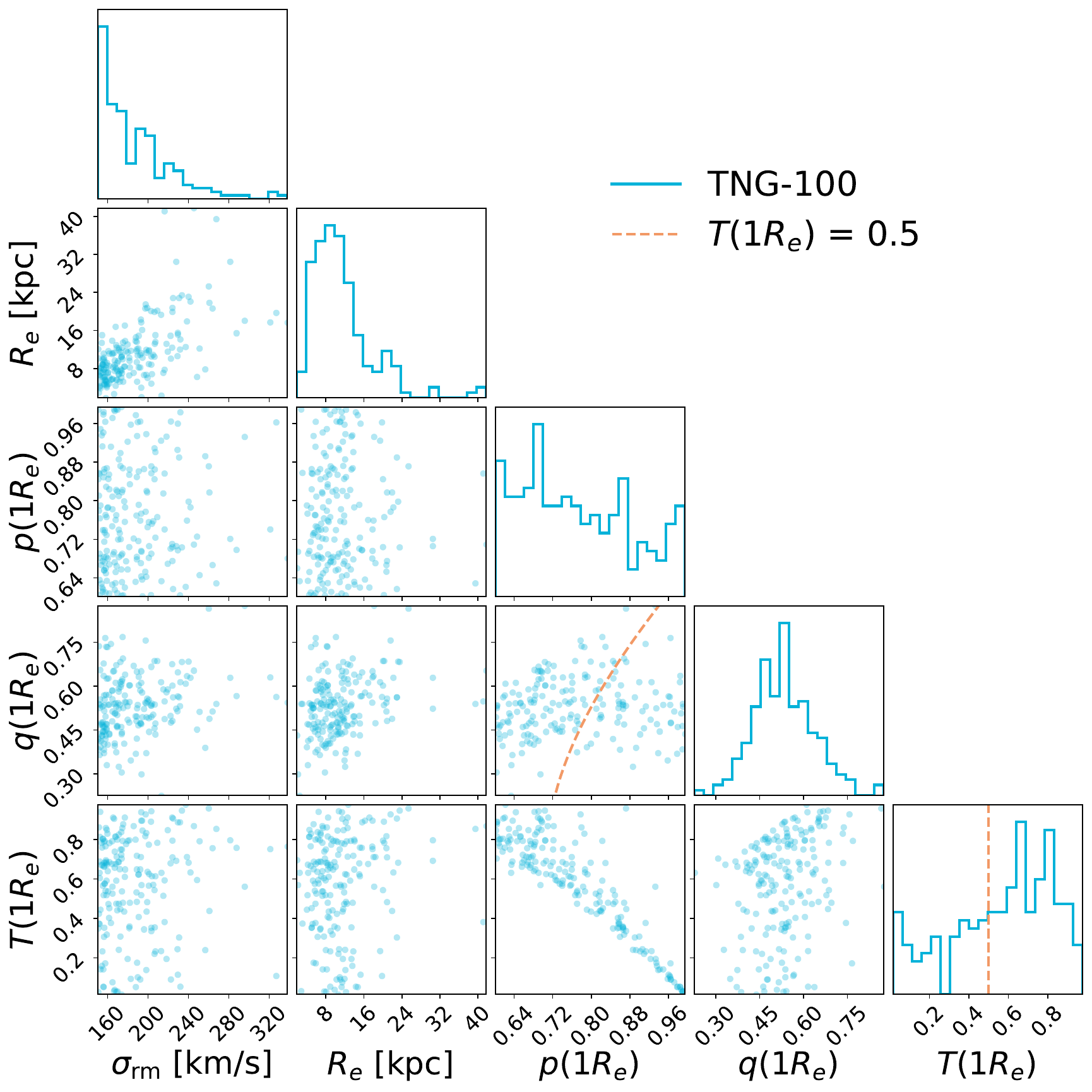}
    \caption{Distribution of the velocity dispersion on a random LoS $\sigma_\mathrm{rm}$,
        the effective radius $R_e$, the intrinsic axis ratios $p$ and $q$ and the triaxiality parameter $T$ for the selected ETG sample from them TNG100 simulation. Dashed line represent $T = 0.5$ in the $p-q$ space and in the 1D histogram of $T$, which we used to separate the full sample into the "oblate" subsample and the "prolate" subsample. }
    \label{fig:triaxiality}
\end{figure}

For the TNG sample, we used $\sigma_\mathrm{rm}$ as an additional mass cut, keeping only galaxies with $\sigma_\mathrm{rm}\geq 150 $ km/s.
This is to construct a more realistic lens sample as lens galaxies are generally the most massive ones with velocity dispersion in the range of 200 to 350 km/s (e.g., \citet{2006Bolton}). We note however, that the distribution of stellar velocity dispersions and ellipticities for the TNG sample is much different than that found for SLACS galaxies. Even after the cut, the TNG sample peaks at the low end of the velocity dispersion distribution and has virtually no galaxies above 240 km/s. The projected ellipticity of the TNG sample under random projection (simulated in Section \ref{subsec:proj_effect_triaxial_tng100}) is also generally larger than that of the SLACS lenses. Therefore, we consider the results based on the TNG sample to be representative of the maximum possible effects, since lens galaxies will generally be rounder and less triaxial, due to selection effect.
We can use the triaxiality parameter,
\begin{equation}\label{eqn:triaxiality_param}
    T = \frac{1 - p^2}{1 - q^2},
\end{equation}
to characterize the intrinsic shape of the sample ETGs.
For prolate ellipsoids, $T = 1$; for oblate ellipsoids, $T = 0$; for triaxial ellipsoids, $0 < T < 1$.
Figure \ref{fig:triaxiality} shows the intrinsic axis ratios and triaxiality distribution of the TNG100 sample.
The full TNG ETG sample is further divided into two subsamples: the oblate sample with $T \leq 0.5$ and the prolate sample with $T > 0.5$, since galaxies of different intrinsic shapes have different projection effects.
Finally, among the TNG100 sample, the oblate subsample has 65 ETGs and the prolate subsample has 126 ETGs.

\subsection{Mass-density, luminosity, and anisotropy assumptions}\label{subsec:lens_gal_model}

For our sample galaxies, we adopted an intrinsic galaxy model composed of a dark matter halo and a stellar component.
The overall density profile is assumed to be a deformed (triaxial or axisymmetric) singular isothermal sphere (SIS), which approximates the population level density profile of massive ETGs \citep{2009Koopmans}.
The Einstein radius of a SIS halo is directly linked to its velocity dispersion by
\begin{equation}\label{eqn:thetaE_sigma_v}
    \theta_{\mathrm{E}}=4 \pi\left(\frac{\sigma_v}{c}\right)^2 \frac{D_{\mathrm{ds}}}{D_{\mathrm{s}}}.
\end{equation}
To better conserve the mass when varying the intrinsic shape of the nonspherical density profiles, we manually add a characteristic "truncation" radius, $r_c$, with $r_c \gg \theta_E$, such that the overall density profile follows
\begin{equation}\label{eqn:trunc_sis}
    \rho(r) \propto \frac{\sigma_v^2}{\left(\frac{r}{r_c}\right)^2\left[1 + \left(\frac{r}{r_c}\right)^2\right]}.
\end{equation}
In this work, $r_c$ is chosen to be 200 times $\theta_E$, far outside the strong lensing regime.  This choice is arbitrary and it does not affect our conclusions. Adopting 100 or 300 would have yielded the same results.
This is because strong lensing measurements do not have constraining power for the density profile far outside the Einstein radius.
We then deformed the spherical iso-density contours into triaxial and axisymmetric ellipsoids, as described in more detail in Section \ref{subsec:projection_formalism} and Section \ref{subsec:projection_form_axisymmetric}.
We chose the normalization factor $Z = (pq)^{1/3}$, such that the spherically averaged density profile is closer to that of a spherical model of the same mass (discussed in detail in Appendix \ref{apdx:norm_factor}).

For the stellar luminosity, we adopted the Jaffe profile \citep{1983Jaffe}, which approximates the de Vaucouleurs profile of elliptical galaxies in projection.
The intrinsic luminosity profile follows
\begin{equation}\label{eqn:jaffe}
    l(r) \propto \left(\frac{r}{r_s}\right)^{-2}\left[1 + \left(\frac{r}{r_s}\right)\right]^{-2},
\end{equation}
where $r_s = R_e / 0.763$ with $R_e$ being the half-light radius. The absolute normalization of the stellar luminosity is arbitrary, since the stellar motion in a galaxy is only dependent on the underlying total mass distribution and the shape of the tracer component.
The predicted LoS velocity dispersion is weighted by the tracer density, whose absolute value is canceled out in the solution of the Jeans equation, expressed in Eqs. (\ref{eqn:axi_jeans_1}-\ref{eqn:axi_jeans_2}).

For the anisotropy in the stellar orbits, we adopted uniformly isotropic orbits, i.e., $\beta=0$ at any radius.
This choice falls in the range of $\beta$ inferred using Schwarzschild's models from the SAURON project \citep{2007Cappellari}, and is shown in Figure \ref{fig:bias_axi_vs_sph} that it does not drastically affect the qualitative results. Without any loss of generality, we could make some simplifying assumptions about our sample of lenses:
\begin{enumerate}
    \item We assumed all the lenses are at $z=0.5$ and all the lensing sources are at $z=1.5$.
    \item For simplicity and to isolate the projection effect, we assumed that the density profile and luminosity profile of the galaxy are perfectly aligned.

\end{enumerate}

\subsection{Kinematics model}\label{subsec:lens_gal_kin_model}

In the kinematics model construction with JAM, we used the galaxy model with an axisymmetrically deformed truncated SIS profile as the total mass-density profile and an axisymmetric Jaffe profile as the stellar luminosity profile.
For simplicity and to isolate the effect of the intrinsic shape and the inclination angle of the lens galaxy, we chose an isotropic anisotropy parameter (i.e., $\beta = 0)$ at all radii,  assuming no black hole mass contribution to the kinematics.
We use the stellar dynamical modeling software\footnote{\url{https://pypi.org/project/jampy/}} \textsc{JamPy} \citep{2008Cappellari,2020Cappellari}, which solves the Jeans equation in Eqs. (\ref{eqn:axi_jeans_1}-\ref{eqn:axi_jeans_2}) and computes the velocity dispersion map in projection.
In Section \ref{subsec:kinematics_mock_data}, we present our mock velocity dispersion data, generated by projecting the 3D kinematics models onto random directions. We then analyze the projection effect and the selection effect.

\section{Projection effects in strong lensing observables}\label{sec:proj_effect_lensing}

In this section, we discuss the projection and selection effect in the strong lensing observables for triaxial lens galaxies.
In Section \ref{subsec:projection_single_triaxial}, we describe the projection effect in the lensing observables for an individual lens galaxy.
In Section \ref{subsec:proj_effect_triaxial_tng100}, we first present the projection effect for a sample of triaxial lens galaxies from the TNG100 simulation and then apply lensing selections to model the selection effect for lens galaxies of different shapes.

\subsection{Projection effect for a single triaxial lens galaxy}\label{subsec:projection_single_triaxial}

\begin{figure}[htbp]
    \centering
    \includegraphics[width=0.95\linewidth]{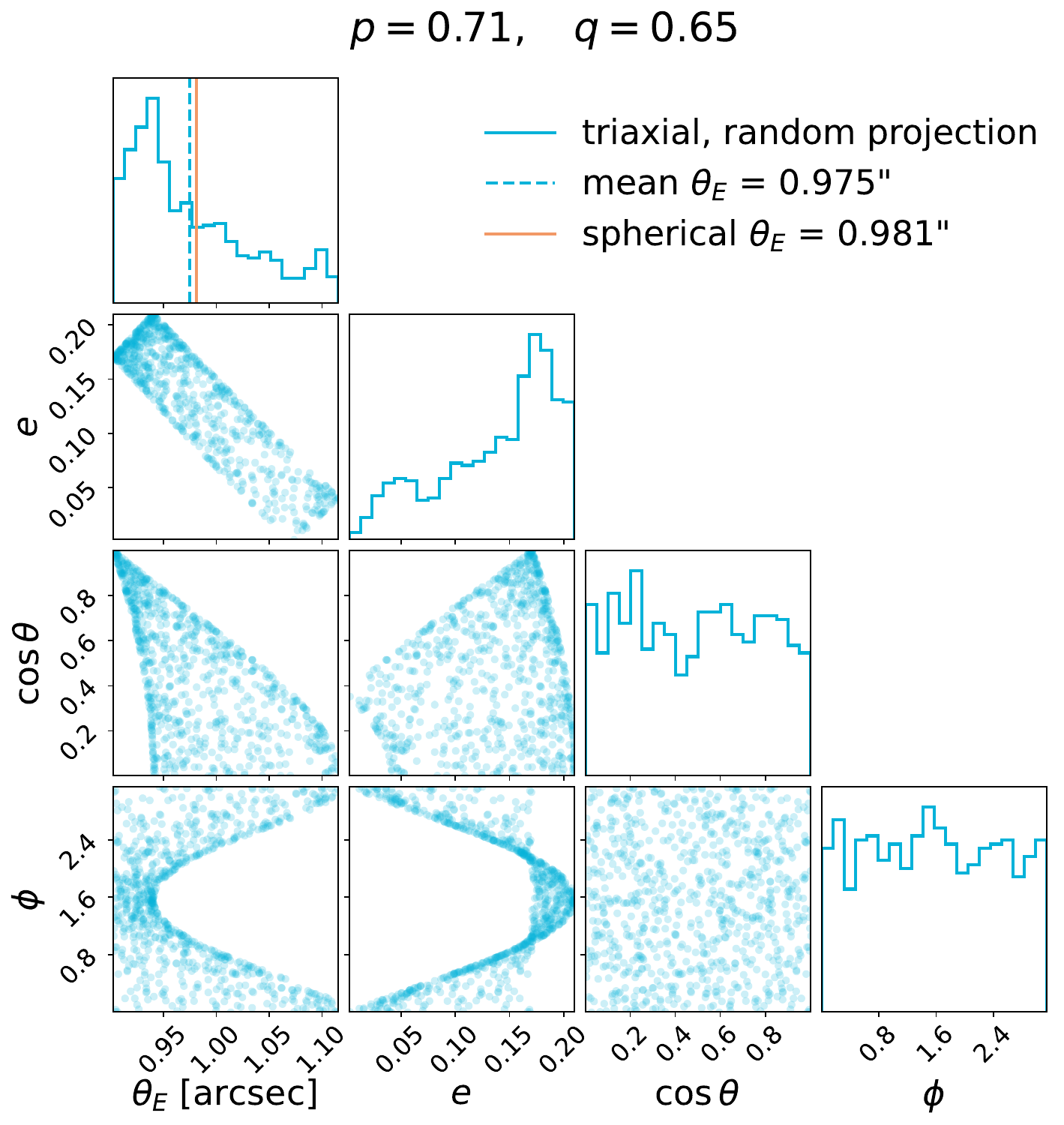}
    \caption{Illustration of the projection effect in the strong lensing observables. We start with 3D density profiles of lens galaxies and project onto random directions 800 times. Dashed blue line marks the mean of $\theta_E$ under random projections. Orange solid line marks the $\theta_E$ of a spherical lens of the same mass. The projection effect in lensing is reflected by the scattering of $\theta_E$ around the spherical value.
    }\label{fig:single_triaxial_projection}
\end{figure}

In this section, we present an example of the projection effect in the strong lensing observables of individual galaxies.
We use the deformed truncated SIS profile in Eq. (\ref{eqn:trunc_sis}) as the total mass-density profile.
We assume the viewing angle $(\theta, \phi)$ is isotropic on a sphere, namely, $\cos \theta \in \mathcal{U}[0, 1]$ and $\phi \in \mathcal{U}[0, \pi]$.
We project the mass-density profile 800 times and calculate the circularized Einstein radius $\theta_E$ and the projected ellipticity $e$.
Figure \ref{fig:single_triaxial_projection} is an example with one triaxial ETG.
Due to the projection effect, the mean of the Einstein radius is slightly different with the Einstein radius of a spherical lens of the same mass.

\subsection{Projection effect for the TNG ETG sample}\label{subsec:proj_effect_triaxial_tng100}

In this section, we model the projection effect of a triaxial lens galaxy sample.
We randomly project the TNG100 ETG sample and calculate the circularized Einstein radius $\theta_E$ from the radial density profile.
To generate more data points, we project each ETG in the sample along 4 random orientations.
We quantify the scatter in the Einstein radius for the sample introduced by the projection effect using the mean of the relative standard deviation (RSD) as
\begin{equation}\label{eqn:quantify_scatter}
    \sum_j^N \frac{\sigma_{\theta_{E, j}}}{\langle \theta_{E, j}\rangle} \frac{1}{N},
\end{equation}
where $j$ represents each individual lens, $N$ is the total number of lenses, $\sigma_{\theta_{E, j}}$ is the standard deviation of the Einstein radius in different projections, and $\langle \theta_{E, j}\rangle$ is the mean Einstein radius in the projections.
The mean scatter for the triaxial TNG100 ETG sample is calculated to be 7.8\%.
The projection effect in the Einstein radius then leads to the uncertainty in the deprojection of the best fit lens mass model for the kinematics prediction.

We define a "bias" term in the inferred SIS-model-equivalent velocity dispersion:
\begin{equation}
    b_\sigma = \frac{\sigma_\mathrm{SIS}}{\sigma_\mathrm{rm}} - 1,
\end{equation}
where $\sigma_\mathrm{SIS}$ is calculated by inverting Eq. (\ref{eqn:thetaE_sigma_v}) with $\theta_E$. If the mass distribution within a lens is spherical, $b_\sigma = 0$. Assuming the SIS mass model for every lens, and taking $\sigma_\mathrm{rm}$ as the observed velocity dispersion, the mass sheet parameter $\lambda$ is simply constrained by (from Eq. (\ref{eqn:sigma_under_mst}))
\begin{equation}
    \lambda = \frac{1}{(b_\sigma + 1)^2}.
\end{equation}

\begin{figure*}[htbp]
    \centering
    \includegraphics[width=0.9\linewidth]{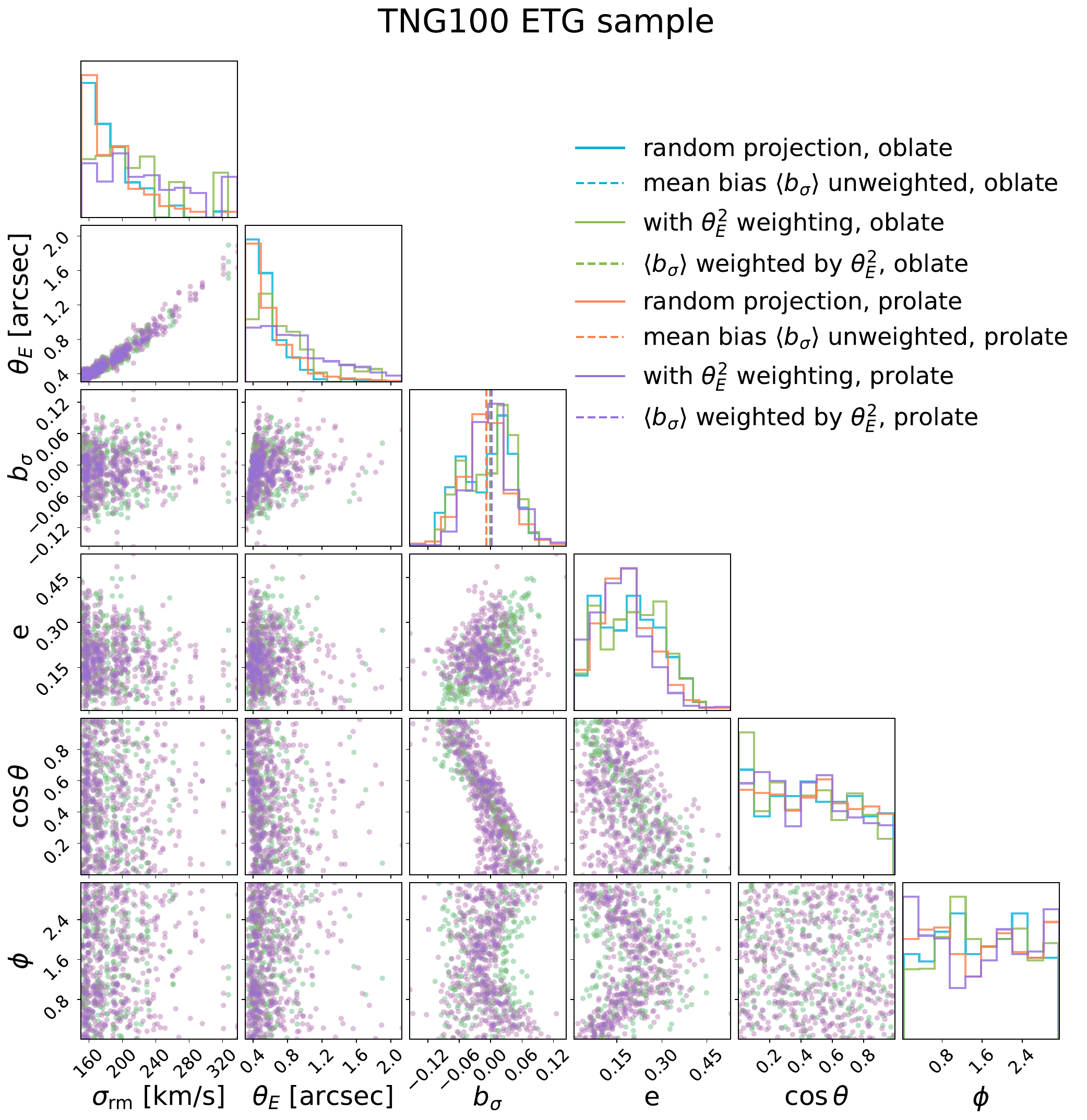}
    \caption{Projection effect of the TNG-100 ETG sample. Each ETG is projected 4 times onto random directions (blue and coral histograms). Each dot in the 2D histograms represents a projection. We apply a lensing cross-section weighting proportional to $\theta_E^2$,  as represented with the green and purple histograms. We model the lensing selection on the projected ellipticity $e$, showing that for the oblate sample, lensing selection favors more elliptical galaxies in projection, while for the prolate sample, the lensing selection favors rounder galaxies. For the oblates, the viewing angle $\theta$ equals the inclination angle $i$ for pure oblates, and thus under lensing selection $\cos\theta$ is also inclined to the lower end, i.e., towards higher inclination angles and consequently higher ellipticities. The "bias" in the kinematics under the assumption of SIS lens models $b_\sigma = \sigma_\mathrm{SIS}/\sigma_\mathrm{rm} - 1$ is labeled with dashed lines. The mean bias $\langle b_\sigma \rangle$ for the oblate sample, indicated with blue dashed lines, is directly under the coral dashed line representing $\langle b_\sigma \rangle$ for the prolate sample and thus invisible from the plot. }
    \label{fig:tng100_random_projection_both}
\end{figure*}

Figure \ref{fig:tng100_random_projection_both} presents the lensing observables under the projection effect for the oblate subsample and the prolate subsample, with each blue dot representing a projection.
Due to the projection effect, the means of the bias, $b_\sigma$, for both samples are overall biased low comparing to the spherical case.

\subsection{Selection effect}

Starting from the lensing observables distribution in Figure \ref{fig:tng100_random_projection_both}, we model the lensing selection effect for the lens galaxies with different shapes.
We apply a lensing cross-section weighting proportional to $\theta_E^2$ to the projected lensing observables \citep{2024Sonnenfeld}. 
The distribution of the lensing selected quantities are shown with green and purple lines and dots in Figure \ref{fig:tng100_random_projection_both}.
For the oblate sample, the lensing selection prefers more elliptical objects in projection, and higher $\theta$, which is the inclination angle $i$ if the galaxy is a pure oblate ($p = 1$).
For the prolate sample, which is relatively rounder, the lensing selection slightly favors less elliptical objects in projection.
The inclination angle $i$ of a pure prolate ($p=q$) has $\cos i = \sin \theta \cos \phi$, and thus the selection in the viewing angles ($\theta, \phi$) is not directly visible from the corner plot.

For both samples, we observe that the bias $b_\sigma$ is relieved with the lensing selection.
This is because the projection effect of each individual galaxy leads to lower stacked surface density around the center, deviating $b_\sigma$ from 0. The lensing selection favors more surface density, and therefore the bias $b_\sigma$ moves in the direction of the spherical case (i.e, 0).

\section{Projection effects in kinematics}\label{sec:kinematics_results}

In Section \ref{sec:proj_effect_lensing}, we use the most simplified kinematics model (i.e., the SIS density profile) to explain how the projection effects in the lensing observables affect the interpretation of the mass sheet parameter $\lambda$.
In this section, we adopt the more flexible JAM for the stellar kinematics modeling and investigate the projection and selection effect.

In Section \ref{subsec:kinematics_mock_data}, we present our mock kinematics data using the axisymmetric synthetic lens sample based on assumed intrinsic mass-density profile and stellar luminosity profile.
We  apply the lensing selection proportional to the lensing cross-section to model the selection effect in the kinematic observables.
In Section \ref{subsec:kin_recovery_sph} we test the recovery of the velocity dispersion for the mock data sample using spherical Jeans modeling.
In Section \ref{subsec:kin_recover_axisym_jam}, we test the recovery of the velocity dispersion for the mock data using axisymmetric JAM.

Throughout this section, we base our discussion of the projection effect on axisymmetric kinematics models, where we use $q$ to represent the intrinsic axis ratio of the ETGs, and $q'$ to represent the projected axis ratio.
For oblates, $\{q, q'\} < 1$; for prolates, $\{q, q'\} > 1$.

\subsection{Mock kinematics data and the lensing selection}\label{subsec:kinematics_mock_data}

\begin{figure}[htbp]
    \centering
    \includegraphics[width=0.98\linewidth]{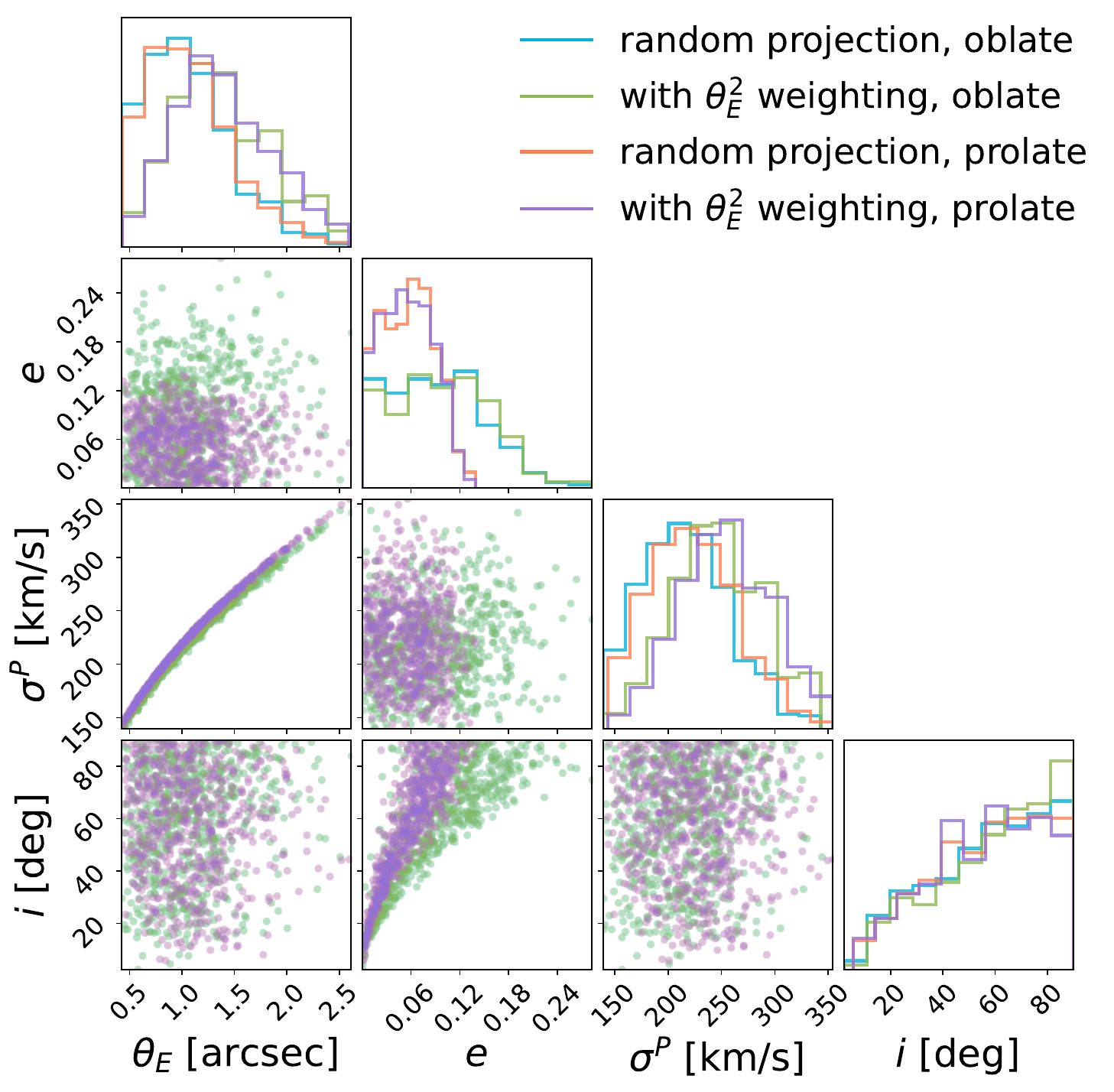}
    \caption{Corner plot of the kinematics and lensing observables for the mock data, generated by projecting the axisymmetric synthetic lens sample assuming random LoS. Each ETG is projected one time. We applied a lensing selection weighted by the cross-section area, $\propto \theta_E^2$, to illustrate the lensing selection effect on the projected ellipticity and the inclination angle. }
    \label{fig:kinematics_observables_separate}
\end{figure}

We use the dynamical modeling software \textsc{JamPy} \citep{2008Cappellari,2020Cappellari} to generate a set of mock kinematics data with the axisymmetric synthetic lens sample.
We start from the intrinsic MGE components of the mass-density and stellar luminosity profiles, project onto random directions, and model the observed velocity dispersion.
The projection formalism of axisymmetric systems under MGE parameterization has been summarized in Section ~\ref{subsec:projection_form_axisymmetric}.
The generation of the mock kinematics data was performed as follows.

\begin{enumerate}
    \item We started with the 3D "deformed, truncated SIS" mass-density profile and the 3D Jaffe stellar luminosity profile of the synthetic lens sample and fit the analytic profile with \texttt{mge\_fit\_1d} from the \textsc{MgeFit} software package\footnote{\url{https://pypi.org/project/mgefit/}} \citep{2002Cappellari} to obtain their intrinsic MGE components.
    \item We projected the intrinsic, axisymmetric MGE components according to Eqs. (\ref{eqn:axis_ratio_proj_oblate} - \ref{eqn:peak_density_proj}) to obtain their projected counterparts.
    \item \textsc{JamPy} was used to de-project the projected MGE components and compute their Jeans solutions. As a result, maps of the projected second velocity moments were generated. We used some simplifying assumptions: no PSF convolution in the projected velocity moments map, no black hole at the galaxy center, isotropic velocity components throughout all radii ($\beta$=0), and perfect alignment between the stellar halo and the total mass profile.

\end{enumerate}

Specifically, for the prolate galaxies, we assumed the symmetry axis lies along the projected major axis and rotate the input coordinate by 90 degrees. These galaxies have orthogonal photometric and kinematics major axes \citep{2017Tsatsi, 2018Li}. Figure \ref{fig:kinematics_observables_separate} shows the distribution of the mock data. The circularized Einstein radius, $\theta_E$, is computed from the projected mass-density MGE components. We further applied a lensing cross-section weighting proportional to $\theta_E^2$ to model the lensing selection effect for the projected ellipticity and the inclination angle. Similar to what we conclude from Figure \ref{fig:tng100_random_projection_both}, the lensing selection prefers higher projected ellipticity and higher inclination angle for the oblate population, while for the prolate population, the lensing selection prefers lower projected ellipticity, and lower inclination angle.

\subsection{Kinematics recovery with spherical JAM}\label{subsec:kin_recovery_sph}

\begin{figure}
    \centering
    \includegraphics[width=\linewidth]{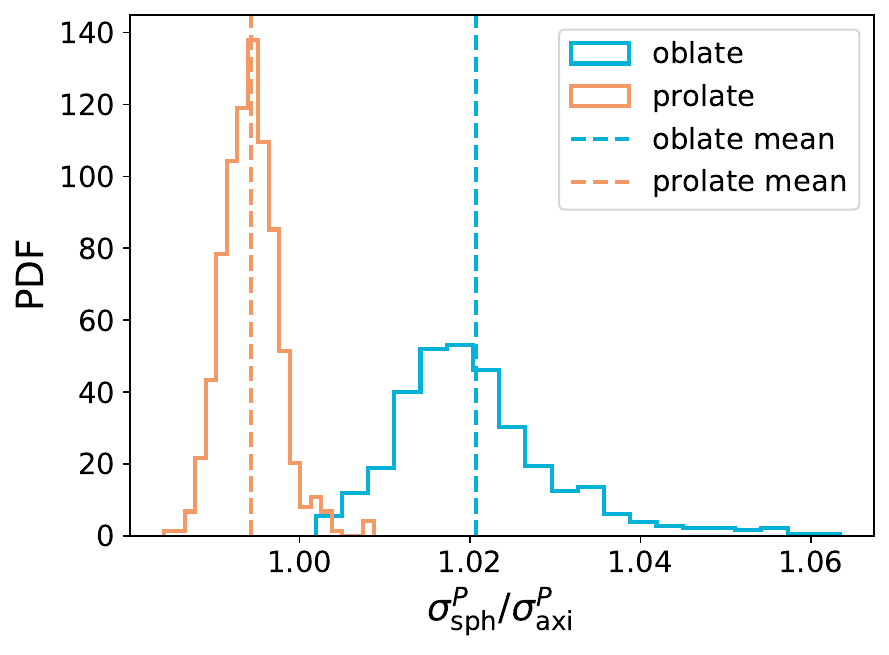}
    \caption{Distribution of the ratio of the velocity dispersion recovered using spherical JAM and generated using axisymmetric JAM, starting from the same 3D models. For oblates, this ratio is larger than unity, while for prolates most values are smaller than unity, showing that the two populations have different bias in the kinematics recovery when spherical model is assumed. The mean of this ratio for oblates is 1.020, and for prolates is 0.994. }

    \label{fig:spherical_kin_recovery}
\end{figure}

\begin{figure}
    \centering
    \includegraphics[width=\linewidth]{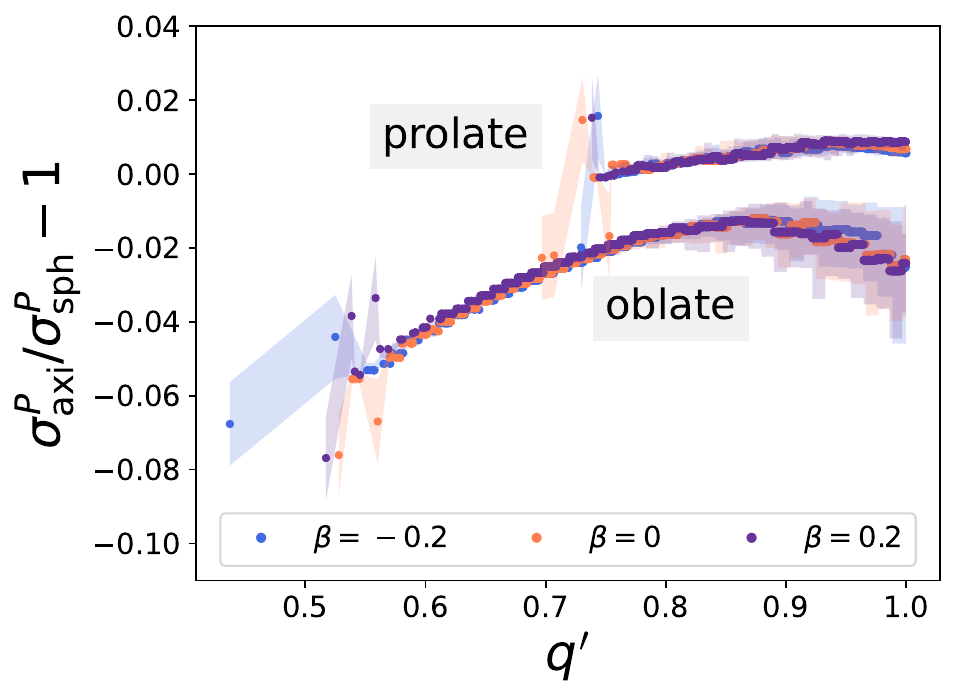}
    \caption{$\sigma_\mathrm{axi}^P/\sigma_\mathrm{sph}^P - 1$ conditioned on the observed axis ratio $q'$ for the oblate intrinsic shape distribution $q\in\mathcal{N}[\mu=0.74, \sigma = 0.08]$ and the prolate $1/q \in\mathcal{N}[\mu=0.84, \sigma = 0.04]$. For this test, the velocity dispersion used to characterize the truncated SIS profile is set to be 200 km/s, and the half-light radius used to characterize the Jaffe profile is set to be 7.5 kpc. The number of draws of the intrinsic axis ratio is 2500, while each data point is projected once assuming isotropic inclination angles. The discreteness of the curve is due to the binning from the 2D histogram of $q'$ vs $\sigma_\mathrm{axi}^P/\sigma_\mathrm{sph}^P - 1$. The solid dots are the median of the conditional probability $P(\sigma_\mathrm{axi}^P/\sigma_\mathrm{sph}^P - 1 \vert q')$, while the shaded area is the $1\sigma$ interval. }
    \label{fig:bias_axi_vs_sph}
\end{figure}

In this section, we focus on the impact from the spherical assumption in the kinematics recovery on the interpretation of the mass-density profile. We start with the mock kinematics data we generated (see Section \ref{subsec:kinematics_mock_data}) and recovered their 3D kinematics model using spherical JAM.
We calculate the spherically averaged MGE for the mass-density profile and the stellar luminosity profile.
The total mass (luminosity) expressed with MGE can be written as
\begin{equation}\label{eqn:total_mge_lum}
    L_k = \sum^N_{k=1} \nu_{0k} (\sigma_k \sqrt{2\pi})^3 q_k,
\end{equation}
and, therefore, the sphericalized MGE has $\sigma_k^\mathrm{sph} = \sqrt{q'}\sigma_k$.
Assuming perfect measurement, we compare the velocity dispersion recovered using spherical JAM and the mock data generated using axisymmetric JAM.
In Figure \ref{fig:spherical_kin_recovery} we show the distribution of the ratio $\sigma^P_\mathrm{sph}/\sigma^P_\mathrm{axi}$ and calculate their mean values.
We find that the recovered velocity dispersion distribution is biased 2\% on average for the oblate sample, and is biased 0.6\% for the prolate sample.
To explain the sign of the percentage error, we refer the reader to Appendix \ref{apdx:norm_factor}, where we compare the stacked axisymmetric surface density profiles and their spherical counterparts.
We use the normalization factor $Z = (pq)^{1/3}$, and as presented in the lower right corner of Figure \ref{fig:sphericalizing_comparison}, the stacked surface density profile of an oblate density profile is slightly lower than that of a spherical profile at small radius, while the prolate case is the opposite.
Therefore, the oblate profile has more mass around the center compared to the spherical case, and therefore contribute more to the velocity dispersion within the effective radius.
The prolate case is the opposite, having less mass around the center and, hence, contributing less to $\sigma^P$.

We specify here that the ratio $\sigma_\mathrm{sph}^P/\sigma_\mathrm{axi}^P$ is independent of the velocity dispersion used to characterize the truncated SIS profile, and thus independent of the overall mass of the galaxy.
The shape of the tracer profile will only slightly impact the value of the $\sigma_\mathrm{sph}^P/\sigma_\mathrm{axi}^P$.
We test using Hernquist profile as stellar tracer, with the same scale radius as the Jaffe profiles.
The change in the value of the ratio is less than 0.5\%. We also test whether the value of $\sigma_\mathrm{sph}^P/\sigma_\mathrm{axi}^P$ is impacted by the value of the scale radius. Assuming Jaffe profile as the tracer, when the scale radius is in the range of [3.28, 26.21] kpc (corresponding to half-light radius of [2.5, 20] kpc), the relative change is less than 0.2\%.
Therefore, we conclude that the choice of the tracer profile and its scale radius are subdominant influencing factors to $\sigma_\mathrm{sph}^P/\sigma_\mathrm{axi}^P$.
The major parameter impacting $\sigma_\mathrm{sph}^P/\sigma_\mathrm{axi}^P$ is the intrinsic axis ratio of the lens population.
We then use a fixed velocity dispersion which characterizes our truncated SIS profile (in terms of mass-density), a fixed scale radius which characterizes the Jaffe profile (stellar tracer), and more realizations of the intrinsic axis ratio drawn from the intrinsic shape prior \citep{2018Li}, to generate a larger sample of modeled $\sigma_\mathrm{sph}^P/\sigma_\mathrm{axi}^P$.

As an example, we drew 2500 intrinsic axis ratios for both the oblate and prolate population and calculate $\sigma_\mathrm{sph}^P/\sigma_\mathrm{axi}^P$, under three choices of the anisotropy parameter $\beta \in \{-0.2, 0, 0.2\}$.
For the purpose of making a correction to the JAM modeling, depending on the observed axis ratio of the lens, we inverted $\sigma_\mathrm{sph}^P/\sigma_\mathrm{axi}^P$ to obtain a multiplicative correction term, $\sigma_\mathrm{axi}^P/\sigma_\mathrm{sph}^P$, and also to condition the distribution of $\sigma_\mathrm{axi}^P/\sigma_\mathrm{sph}^P$ on the observed axis ratio. The distribution of $\sigma_\mathrm{axi}^P/\sigma_\mathrm{sph}^P - 1$ (the -1 is for visualization) versus the projected axis ratio $q'$ is shown in Figure \ref{fig:bias_axi_vs_sph}.
While $\sigma_\mathrm{axi}^P/\sigma_\mathrm{sph}^P - 1$ for the prolate population is relatively flat for different $q'$, the oblate curve shows a tight correlation between $\sigma_\mathrm{axi}^P/\sigma_\mathrm{sph}^P - 1$ and $q'$.
Both features make it possible to estimate the systematic uncertainty in the model prediction of the velocity dispersion under spherical assumption.
The prolate curve has a smaller dispersion since the population is intrinsically rounder. 
For the oblates, the residual uncertainty (the 1$\sigma$ interval) increases at both ends of $q'$ due to the sparsity of data points. 
For the more elliptical end ($q' \sim0.5$), the lack of data points is due to the intrinsic shape distribution of the population, since the most elliptical projection has $q' = q$, $i = 90^\circ$.
For the rounder end, the lack of data points has two causes: the intrinsic shape distribution and the distribution of isotropic inclination angles.
The isotropic inclination angle satisfies $\cos i \in \mathcal{U}[0, 1]$, and therefore galaxies tend to have higher inclination angles if no selection effect is considered, resulting in less round galaxies in projection.
To sum up, assuming an axisymmetric galaxy population, one can precisely correct for the spherical Jeans modeling for galaxies given solely the intrinsic axis ratio distribution of the population.

\begin{figure}
    \centering
    \includegraphics[width=\linewidth]{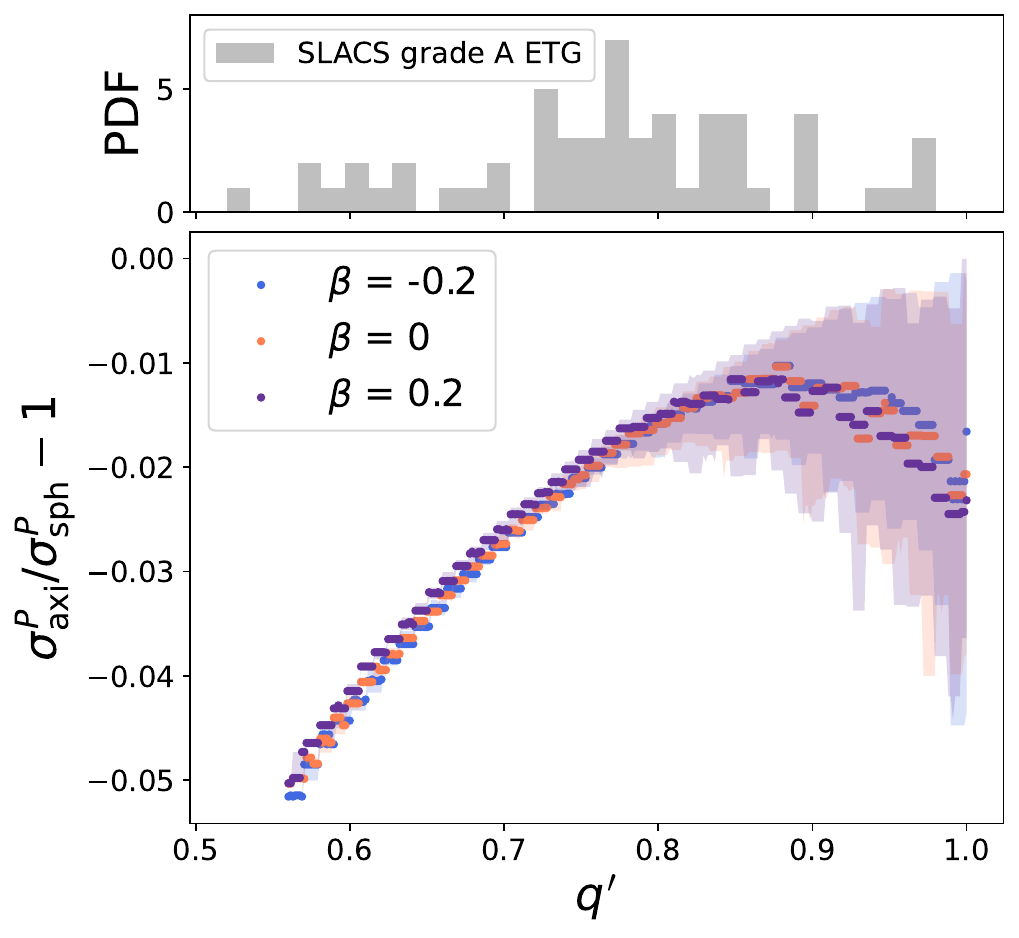}
    \caption{Kinematics correction term $\sigma_\mathrm{axi}^P/\sigma_\mathrm{sph}^P - 1$ versus the projected axis ratio $q'$ as a result of combining the two curves in Figure \ref{fig:bias_axi_vs_sph} (lower panel). We use the misalignment angle between the kinematics major axis and the photometric major axis of the SLACS lenses measured in \citet{2024Knabel}, and a simple model for the fraction of oblates in a population (see Appendix \ref{apdx:oblate_fraction}), to determine the weight used to combine the two curves.  The solid dots in the lower panel are the median of the conditional probability $P(\sigma_\mathrm{axi}^P/\sigma_\mathrm{sph}^P - 1 \vert q')$, while the shaded area is the $1\sigma$ interval. The upper panel is to illustrate the distribution of the projected axis ratios of the early-type SLACS lenses from \citet{2008Bolton}. Averaging over all these SLACS lenses, and under the choice of $\beta = 0$, we obtain the median correction to be $(\sigma_\mathrm{axi}^P/\sigma_\mathrm{sph}^P - 1)_\mathrm{SLACS} = {-0.016}^{+0.009}_{-0.007}$.}
    \label{fig:bias_axi_vs_sph_combined}
\end{figure}

Furthermore, we can combine the correction $\sigma_\mathrm{axi}^P/\sigma_\mathrm{sph}^P - 1$ for the oblate and prolate population assuming that the real lens population is a mix of prolates and oblates.
We use the kinematics misalignment angle measured from the spatially resolved spectra of a subsample of the SLACS lenses from \citet{2024Knabel} to determine the fraction of oblates in our population, and marginalize over the fraction of oblates to combine the two $\sigma_\mathrm{axi}^P/\sigma_\mathrm{sph}^P - 1$ curves for the oblate and prolate samples. We refer the readers to Appendix \ref{apdx:oblate_fraction} for a detailed description of the determination of the fraction of oblates in our population.
The combined correction term is shown in Figure \ref{fig:bias_axi_vs_sph_combined} in the lower panel.
For a lens population with unknown intrinsic shape distribution, the combined $\sigma_\mathrm{axi}^P/\sigma_\mathrm{sph}^P - 1$ versus $q'$ curve can be used to apply correction to the velocity dispersion modeled with spherical JAM.
The uncertainty in the correction term $\sigma_\mathrm{axi}^P/\sigma_\mathrm{sph}^P - 1$ is propagated to the velocity dispersion after correction (effectively $\sigma_\mathrm{axi}^P$) as a residual uncertainty. In our exercise, depending on the choice of the anisotropy parameter and the projected ellipticity, the residual uncertainty is in the range of 0\% - 2.2\%, resulting in an uncertainty of 0\% - 4.4\% in the inferred mass sheet parameter $\lambda$, and therefore $H_0$. 
In the upper panel of Figure \ref{fig:bias_axi_vs_sph_combined}, we plot the histogram of the projected axis ratios of the early-type SLACS lenses from \citet{2008Bolton}, to illustrate the range of $q'$ from observed data. 
Marginalizing over all the $q'$ of the SLACS ETG plotted in Figure \ref{fig:bias_axi_vs_sph_combined}, and under the choice of $\beta = 0$, we obtain the median correction term $(\sigma_\mathrm{axi}^P/\sigma_\mathrm{sph}^P - 1)_\mathrm{SLACS} = {-0.016}^{+0.009}_{-0.007}$. 
Translating the median correction $(\sigma_\mathrm{axi}^P/\sigma_\mathrm{sph}^P - 1)_\mathrm{SLACS}$ into the median "bias" $(\sigma_\mathrm{sph}^P/\sigma_\mathrm{axi}^P - 1)_\mathrm{SLACS}$, we obtain $(\sigma_\mathrm{sph}^P/\sigma_\mathrm{axi}^P - 1)_\mathrm{SLACS} = {0.017}^{+0.007}_{-0.009}$. 
In other words, in the absence of spatially resolved kinematics, treating the lenses as spherical in the JAM modeling could bias the modeled velocity dispersion up by 1.7\%, if not corrected for. 
After correction, the random uncertainty in $\sigma^P$ is 0.8\%.

We conclude that the relative uncertainty introduced by using spherical models rather than axisymmetric models is non-negligible in the population analysis of the kinematics of ETG with JAM.
Therefore, axisymmetric dynamical models should be used to accurately recover the mass distribution of ETGs.

\subsection{Kinematics recovery using axisymmetric JAM}\label{subsec:kin_recover_axisym_jam}

In this section, we discuss the effect of the intrinsic shape of the lens galaxies in the recovery of their kinematics.
The question we want to address is how to best model the velocity dispersion of the lens galaxies under axial symmetry, given their observed ellipticities, the stellar luminosity profiles, the mass-density profiles from lens mass modeling, and an assumed anisotropy profile.
We assume that there is no information from spatially resolved spectra, and therefore the inclination angle cannot be directly obtained for an individual lens galaxy.
We investigate whether the projected kinematics of a sample of galaxies can be recovered, either with or without the prior knowledge of the intrinsic shape of the population. These two approaches are described in Section \ref{subsubsec:kin_recover_axi_Pie}.

We describe how we used the mock data from Section \ref{subsec:kinematics_mock_data}, where we project the axisymmetric synthetic lens sample and recorded the projected ellipticities, the MGE of the projected stellar luminosity profile, and the MGE of the mass-density profile.
We used these observables as input for axisymmetric JAM and compare the recovered velocity dispersion distribution $\sigma^P$ with the mock data. The results are presented in Section \ref{subsubsec:kin_axisym_recovery_results}.

\subsubsection{Deproject kinematics with and without the intrinsic shape information of the population}\label{subsubsec:kin_recover_axi_Pie}

\begin{figure*}
    \centering
    \includegraphics[width=.9\linewidth]{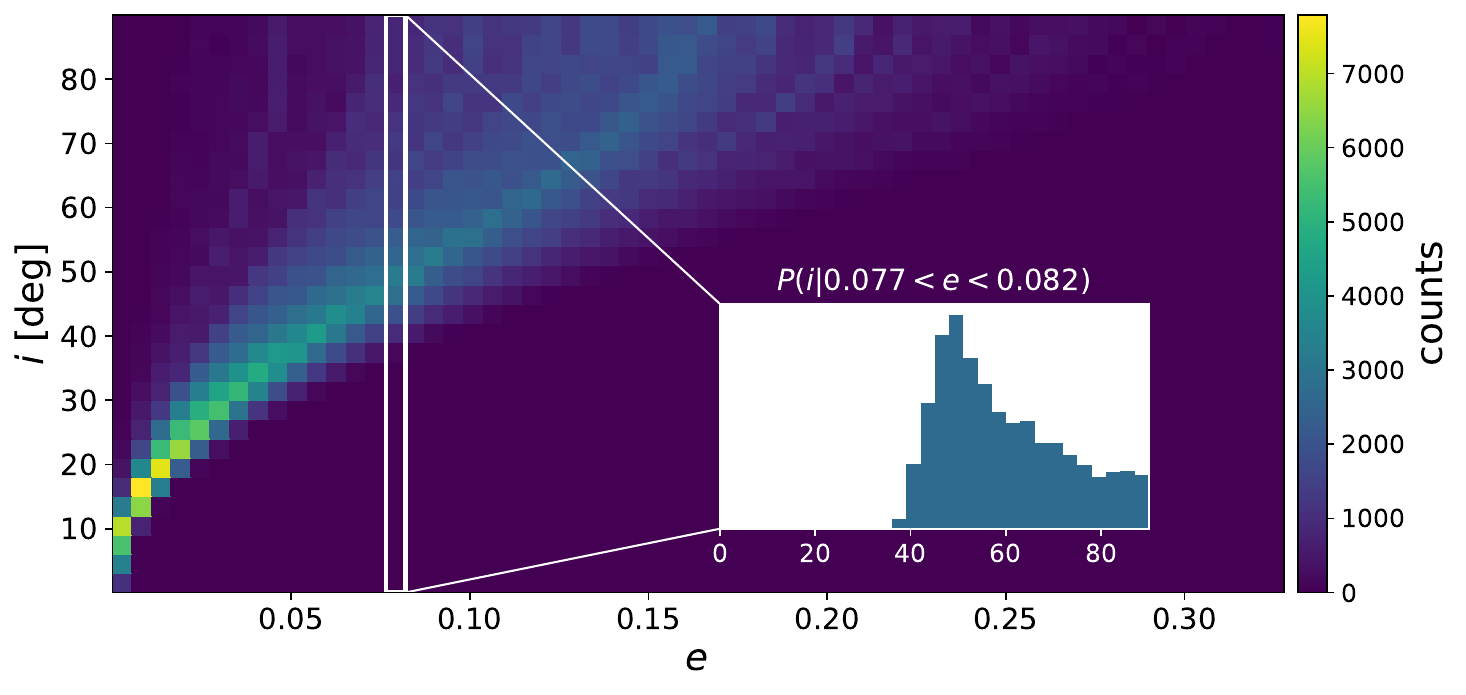}
    \caption{2D histogram of the inclination angle $i$ versus the projected ellipticity, $e$, modeled for the oblate synthetic sample assuming isotropic inclination angles. In the lower right corner is a slice of $P(i\vert e)$ at $e = 0.08$ as an illustration of the shape of the inclination angle distribution.}
    \label{fig:2dhist_i_vs_e}
\end{figure*}

We first describe a framework of proposing a distribution of the inclination angle based on the projected ellipticity and the intrinsic axis ratios of the lens galaxies.

The inclination angle is not isotropic for a galaxy with projected ellipticity, $e$. According to Eq. (\ref{eqn:axis_ratio_proj_oblate}), there is a lower limit of the inclination angle, $i_\mathrm{min} = \arccos{q'}$, where $q'$ is the apparent axis ratio of the projected elliptical isophote. At the minimum inclination angle $i_\mathrm{min}$, the intrinsic axis ratio, $q$, deprojected with apparent axis ratio $q'$ is 0 (i.e., the intrinsic shape of the galaxy's mass density-and-stellar luminosity distribution is extremely elongated). To avoid such an unphysical situation, we set a lower limit of the deprojected intrinsic axis ratio to be $q_\mathrm{min} = 0.05$, noting that this is a very wide prior. The inclination angle distribution for each individual lens galaxy then has a lower bound, $i_\mathrm{min}$, which satisfies
\begin{equation}\label{eqn:minimum_inc}
    q^2_\mathrm{min}\sin^2 i_\mathrm{min} + \cos^2 i_\mathrm{min} = q'^2.
\end{equation}

The distribution of the inclination angle for the sample satisfies
\begin{equation}
    P(i) = \int \mathrm{d}e P(i\vert e) P(e),
\end{equation}
where $P(e)$ is the distribution of the projected ellipticity and
$P(i\vert e)$ is the conditional probability of the inclination angle $i$ given a projected ellipticity, $e$.
Here, $P(e)$ is an observable from the data, while $i$ is not.
We modeled $P(i\vert e)$ with the intrinsic axis ratio distribution of the sample, assuming that the inclination angle is isotropic.
We created a 2D histogram of the inclination angle $i$ versus the projected ellipticity, $e$, with a large number of sample points by drawing from the intrinsic shape distribution.
Then, for each $q$, we used  isotropic inclination angles project for 1000 times using Eq. (\ref{eqn:axis_ratio_proj_oblate}) and calculated the projected ellipticity $e = (1-q') / (1+q')$.
The 2D histogram of the inclination angle $i$ versus the projected ellipticity, $e$, is provided in Figure \ref{fig:2dhist_i_vs_e}.
In the dimension of the projected ellipticity, we use 30 bins.
For the deprojection of each individual galaxy with ellipticity $e$, we take a slice $P(i\vert e \in e^i)$, where $e^i$ is the ellipticity bin.
An example of $P(i\vert e=0.08)$ is included in Figure \ref{fig:2dhist_i_vs_e}.
We then drew the inclination angle from $P(i\vert e)$ and modeled its projected velocity dispersion with axisymmetric JAM, using the projected MGE for the stellar luminosity profile and the mass-density profile. Using the conditional probability $P(i\vert e)$, we recovered the inclination angle distribution for the galaxy sample.
We then used the recovered inclination angle to recover the velocity dispersion distribution.

Another approach to recover the inclination angles is to assume galaxies are randomly oriented and the intrinsic shape is only constrained by their projected ellipticities. In other words, in such a case, we would not include any intrinsic shape prior of the population in the inclination angle recovery.
Then, for each ETG, we draw uniformly from the random inclination angle distribution, and reject the inclination angles which are physically forbidden by the projected ellipticity, according to Eq. (\ref{eqn:minimum_inc}).

\subsubsection{Results of axisymmetric kinematics recovery with and without intrinsic shape prior}\label{subsubsec:kin_axisym_recovery_results}

\begin{figure*}
    \centering
    \includegraphics[width=\linewidth]{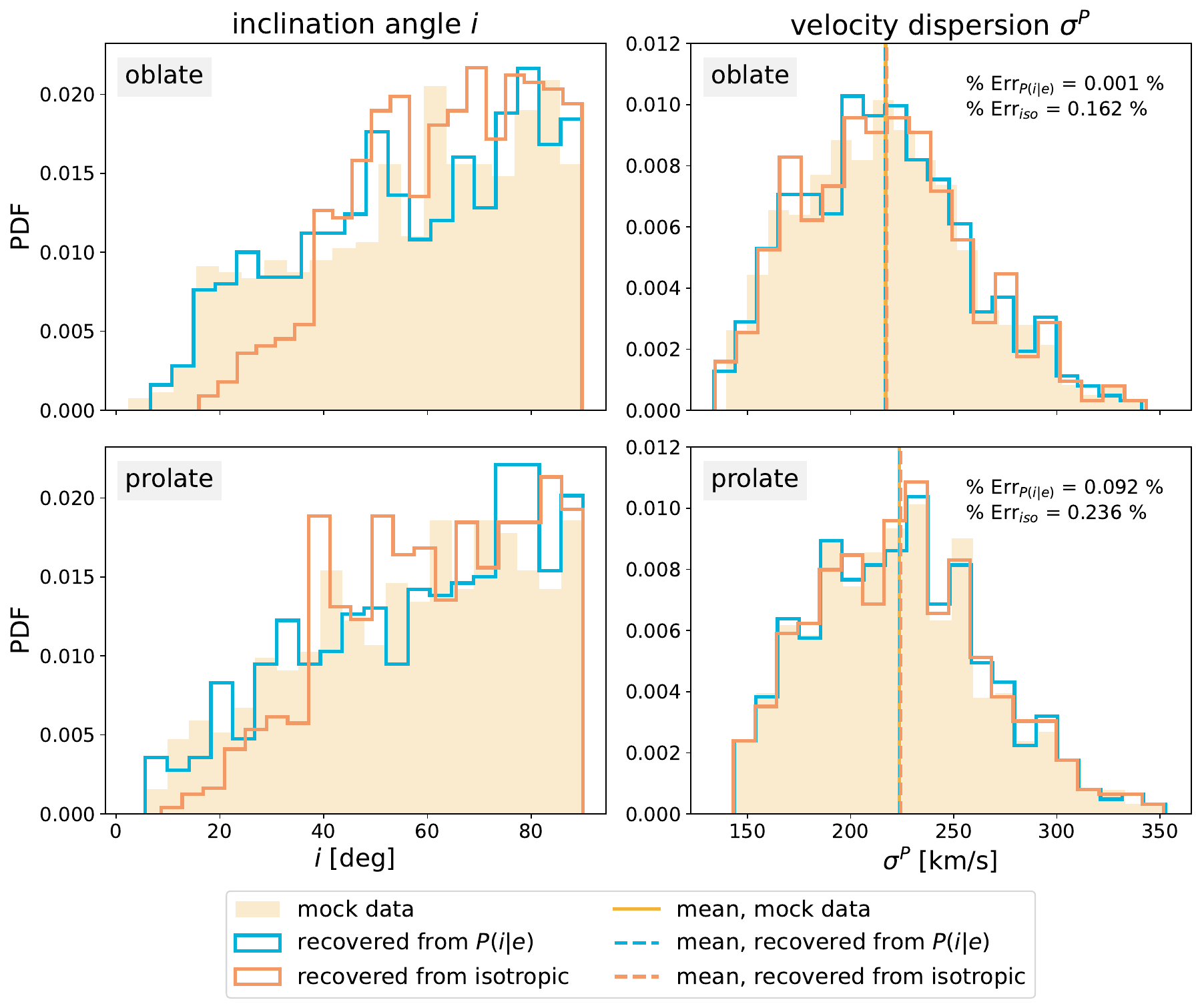}
    \caption{Distributions of the recovered inclination angles and the velocity dispersions using axisymmetric JAM, with (denoted by "recovered from $P(i\vert e)$") and without intrinsic shape prior (denoted by "recovered from isotropic").
       The upper and lower rows show the results for the oblate and the prolate sample, respectively.
        The inclination angle recovered with the intrinsic shape prior has a similar distribution as the inclination angle from the mock data which are sampled isotropically on a sphere, while the inclination angle recovered without the intrinsic shape prior are overall larger than the mock data.
        However, the deviation in the recovered inclination angle does not affect the recovered velocity distribution.
        For both the oblate and the prolate sample, the relative error in the mean of the recovered velocity dispersion is less than 0.24\%.
        This result shows that for a ETG population with the intrinsic shape distribution similar to the one in \citet{2018Li}, the axisymmetric JAM kinematics recovery can be accurate even without the intrinsic shape prior. }
    \label{fig:kin_recovery_axisym}
\end{figure*}

In Section \ref{subsubsec:kin_recover_axi_Pie} we described two methods to recover the inclination angle for a population of ETGs in the absence of spatially resolved spectra.
We apply both approaches to the mock data to obtain a comparison between the recovered inclination angle distribution and the velocity dispersion distribution.
The results are shown in Figure \ref{fig:kin_recovery_axisym}, in which we calculate the relative error between the mean of the recovered velocity dispersion and the mock data.
When the intrinsic shape prior is included (denoted in Figure \ref{fig:kin_recovery_axisym} by "recovered from $P(i\vert e)$") , the distribution of the recovered inclination angle is very similar to the inclination angle distribution from the mock data, which is sampled isotropically on a sphere.
The inclination angle recovered without the intrinsic shape prior (denoted in Figure \ref{fig:kin_recovery_axisym} as "recovered from isotropic") are overall larger than the mock data for both the oblate and prolate sample, due to the fact that the inclination angle at any projected ellipticity has a distribution that is the shape of the isotropic one but with a lower cutoff.
Meanwhile, the actual conditional probability $P(i\vert e)$, as illustrated in Figure \ref{fig:2dhist_i_vs_e}, is peaked at a smaller value.
Having understood the bias in the recovered inclination angle, we find that the recovered velocity dispersion is not impacted.
The relative error between the mean of the recovered velocity dispersion and the mock data is less than 0.24\% with or without the inclination angle prior.

We conclude that for a ETG population with an underlying intrinsic shape distribution of $q\in\mathcal{N}[\mu=0.74, \sigma=0.08]$ for oblates and $1/q \in \mathcal{N}[\mu=0.84, \sigma=0.04]$ for prolates, the kinematic recovery using axisymmetric JAM is accurate with or without knowledge of the intrinsic shape distribution.
However, we note that this conclusion is impacted by the choice of the intrinsic shape distribution, and therefore our conclusion is applicable for an ETG population which is overall less elliptical than our sample.

\section{Triaxiality in kinematics}\label{sec:triaxiality_kinematics}

In Section \ref{sec:kinematics_results}, we discuss the projection and selection effect in the kinematics assuming axisymmetric kinematics models; namely, galaxies, if they are nonspherical, are either oblate or prolate.
Real galaxies, however, can be triaxial in both the mass distribution and the stellar tracer distribution.
Therefore, when we perform axisymmetric kinematic recovery using the observed stellar tracer profile, we ought to bear in mind that the ellipticity, the amplitude, and the shape of the tracer profile is actually the projection of a triaxial ellipsoid.
In this section, we assess the uncertainty in the kinematics recovery from assuming axisymmetric mass and tracer distributions for systems with axisymmetric mass distribution and triaxial tracer distribution, noting that the kinematic model of a system with both triaxial mass and tracer distribution is beyond the scope of this work and needs further mathematical derivation or simulation.
We then compare the velocity dispersion recovered using the ellipticity projected by triaxial tracer profiles to the velocity dispersion modeled using axisymmetric mass and tracer distributions.
Practically, we project the triaxial and the axisymmetric galaxy light profiles of the same total luminosity under the same inclination angle.
For the axisymmetric model, we record the projected velocity dispersion.
For the triaxial model, we record the projected ellipticity and the MGE components, then use them to recover the projected velocity dispersion assuming axisymmetric models.

\subsection{Axisymmetric kinematics models for the TNG sample}

In Section \ref{subsubsec:tng_intro}, we introduce a triaxial ETG catalog from the TNG100 simulation.
In the present section, we  use them to generate approximate axisymmetric JAM models and an ellipticity distribution, $e$, which matches the random projection of the triaxial sample. We start by describing how we approximate the triaxial ETG with axisymmetric ones. For the oblate sample, which has $T \leq 0.5$, we take $p \rightarrow 1$, and $q \rightarrow q(1 + p)/2$.
For the prolate sample with $T > 0.5$, we take $p = q \rightarrow (p + q) / 2$.
Then, we project them randomly, each 10 times to increase data size, and record the projected ellipticity, the projected velocity dispersion, the MGE parameters, and the viewing angles $(\theta, \phi)$.

\subsection{Axisymmetric kinematics recovery for a triaxial population}\label{subsec:kin_recovery_triaxial_ellipticity}

\begin{figure*}
    \centering
    \includegraphics[width=0.99\linewidth]{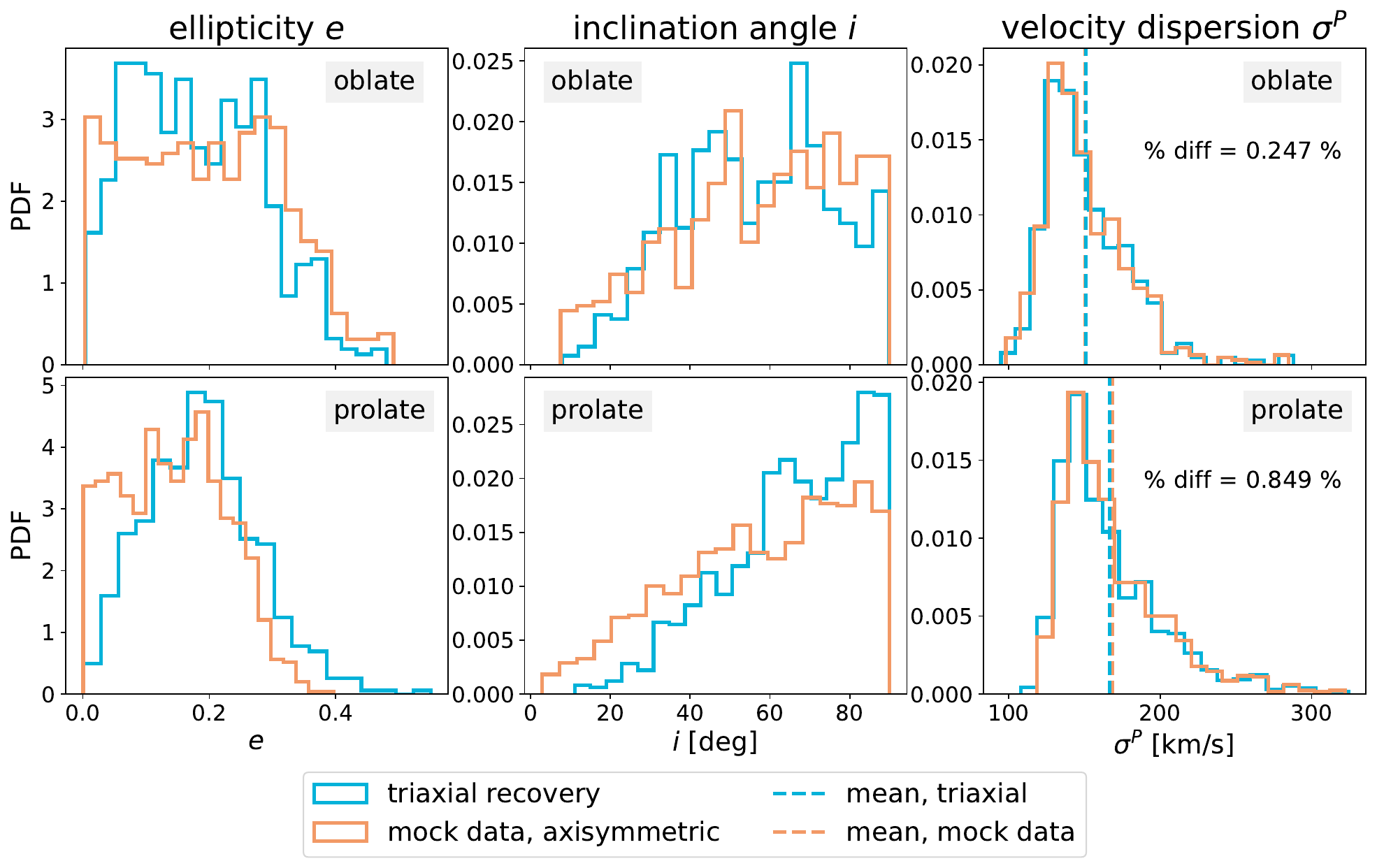}
    \caption{Ellipticity, recovered inclination angle and recovered velocity dispersion for triaxial tracer + axisymmetric mass models and comparison with the axisymmetric tracer + mass models for the triaxial TNG-100 ETG sample.
        The inclination angle distribution is recovered using the conditional probability $P(i\vert e)$, and the velocity dispersion is computed with axisymmetric JAM. We conclude that under the assumption of axisymmetric kinematics models, when the ellipticity is contributed by a triaxial tracer component, the kinematics recovery for the sample is unbiased. }
    \label{fig:triaxial_ellipticity}
\end{figure*}

Using the viewing angles $(\theta, \phi)$, we return to the triaxial models to calculate the projected ellipticity $e^\mathrm{tri}$ using Eq. (\ref{eqn:def_projcted_q}). We also modify the MGEs to preserve the total mass and the total luminosity.
Following Eq. (\ref{eqn:total_mge_lum}), the MGE for the triaxial ellipticities has $\sigma_k^\mathrm{tri} = \sqrt{q'/q_\mathrm{tri}'} \sigma_k$, where $q_\mathrm{tri}'$ is the projected axis ratio corresponding to $e^\mathrm{tri}$.
We then use the modified MGE and the triaxial ellipticities as input for JAM to recover the velocity dispersion $\sigma^P$, and compare with the observables of the axisymmetric models.
We use $P(i\vert e)$ to recover the velocity dispersion for the sample, based on the conclusion in Section \ref{subsubsec:kin_recover_axi_Pie}.
The results are shown in Figure \ref{fig:triaxial_ellipticity}.
The percentage difference between the mean of the recovered velocity dispersion and the mock data is smaller than 0.85\%, from which we validate that the kinematics recovery is unbiased, when the projected ellipticity is contributed by the intrinsic triaxiality of the sample.

We further derive a percentage difference in the mean of the velocity dispersions weighted by the ellipticity of the SLACS early-type grade "A" lenses, as the TNG sample we use is more elliptical and thus the relative difference is not typical for observed lenses. We use the ratio between the density function of the ellipticity of the SLACS lenses and the triaxial ellipticities $w(e) = P^\mathrm{SLACS}(e) / P^\mathrm{tri}(e)$ as a weight term, to calculate the mean of the velocity dispersions of our axisymmetric galaxy models and the mean of the velocity dispersions recovered using triaxial ellipticities. The percentage difference for the velocity dispersion is 0.053\% for the oblate TNG sample and 0.17\% for the prolate TNG sample. We note that the underlying assumption here is  that the SLACS lenses have the same triaxial intrinsic shape distribution as the TNG ETGs, which we cannot yet verify. However, our TNG ETG sample is more triaxial than nearby population of observed ETGs \citep{2018Li}. Therefore, we can conclude that axisymmetric kinematic models can aptly recover the velocity dispersion for a galaxy with axisymmetric mass distribution and triaxial stellar light distribution.

\section{Conclusions}\label{sec:conclusion}

The intrinsic shapes of strong lensing galaxies lead to projection effects in both lensing and stellar kinematics observables. The latter is used to break the mass-sheet degeneracy (MSD), which is crucial for accurate measurements of the Hubble constant with time-delay cosmography.
Since the projection effect affects the lensing convergence, it can lead to selection effects; for example, when the probability of observing a lens is approximately proportional to the lensing cross-section, $\propto \theta_E^2$.
In this work, we quantitatively investigated the projection and selection effects introduced by the intrinsic shape of lens galaxies, using an axisymmetric synthetic lens sample similar to the SLACS lenses and a selected triaxial ETG sample from the TNG100 simulation.

In this work, we  present numerical simulations of the projection effect for the lensing observables for individual triaxial galaxies.
Based on the projection effect and by comparing to spherical lens models, the lensing observables of nonspherical lenses end up scattered around that of the spherical lens of the same mass for 7.8\%.
We were able to quantitatively analyze the selection effect introduced by the projection effect of a triaxial galaxy sample by forward-modeling the TNG100 sample and applying selection function in the form of a lensing cross-section weighting.
We investigated the selection effects in the projected ellipticity of the samples as a function of intrinsic shape distributions, finding that the more oblate galaxies prefer higher projected ellipticity, and the more prolate galaxies prefer lower projected ellipticity.
We demonstrated that the lensing mass estimates can be biased when constrained with nonspherical parameterized mass-density profiles, if the projection effects are not accounted for.

We discuss the projection and selection effect in the stellar kinematics of lens galaxies, based on our use of the spherically aligned axisymmetric JAM to construct stellar kinematic models for our axisymmetric synthetic lens sample as a set of mock data. Using the set of mock data, we model and illustrate the projection effect in the luminosity-weighted velocity dispersion on both individual galaxies and the galaxy sample.

We quantified the bias in the deprojection of the stellar kinematics under different assumptions on the intrinsic shape with the mock data.
One of our main conclusions is that assuming spherical JAM solution biases velocity dispersion modeled with the best-fit lens mass model of imaging data.
For our oblate subsample, the mean bias is 2\%. For our prolate sample, the mean bias is -0.6\%.
This bias is then doubled in the inference of the MSD parameter $\lambda$ and $H_0$.
Moreover, we find that this bias is a function of the projected axis ratio, which enables us to construct a correction term of $\sigma_\mathrm{axi}^P/\sigma_\mathrm{sph}^P - 1$ for the velocity dispersion modeled spherical JAM. The residual uncertainty in the correction term is in the range of 0-2.2\%. This residual uncertainty contributes to the overall uncertainty on the mass sheet parameter, $\lambda$, and, therefore, $H_0$.
As a more specific example, we calculated the median of the correction term using the projected axis ratio of the early-type SLACS lenses from \citet{2008Bolton} to be $(\sigma_\mathrm{axi}^P/\sigma_\mathrm{sph}^P - 1)_\mathrm{SLACS} = {-0.016}^{+0.009}_{-0.007}$, resulting in a residual uncertainty in the corrected velocity dispersion of 0.8\%.

Conversely, using axisymmetric models to deproject the kinematics, we can accurately recover the 3D kinematic models. We presented an upper limit on the overall ellipticity of the ETG sample at which the kinematics recovery using axisymmetric JAM will not be impacted by the inclusion of the intrinsic shape prior.

We also find that assuming axisymmetric JAM models for systems with triaxial stellar light profiles and axisymmetric mass distributions can accurately retrieve the velocity dispersion distribution, using the TNG100 ETG sample. Since the TNG sample is generally more elliptical than lensing galaxies, the conclusion applies to observed lenses, which have an underlying rounder intrinsic shape distribution.

In summary, we developed the machinery for forward-modeling the projection and selection effect for the lensing and kinematics observables of nonspherical galaxies that act as strong lenses.
We showed that the projection effects have
non-negligible impact on the lensing and kinematic observables, finding that an accurate analysis of lensing and kinematic data requires accurate deprojections.
Furthermore, when looking at a population of deflectors, treating the deflectors with spherical assumption does not average out biases on the population level.
Hence, to accurately recover $H_0$ when using a lensing+kinematics analysis to break the MSD requires an explicit treatment of the projection effects on lensing and kinematics data in the analysis.

The framework and quantitative characterization of the lensing+kinematic selection and the axisymmetric versus spherical kinematic treatment presented in this work have been incorporated in the TDCOSMO2025 cosmological results paper \citep{TDCOSMO_milestone} and will also be included in forthcoming time-delay cosmography analysis (e.g., Paic et al., submitted to A\&A).

\begin{acknowledgements}

    The software/packages we use in this project are: \href{https://github.com/huangxy256/deproject.git}{\textsc{deproject}}, \href{https://github.com/lenstronomy/lenstronomy}{\textsc{Lenstronomy}} \citep{lenstronomy1, lenstronomy2}, \href{https://pypi.org/project/jampy/#toc-entry-1}{\textsc{JamPy}} \citep{2020Cappellari}, \href{https://pypi.org/project/mgefit/#toc-entry-1}{\textsc{MgeFit}} \citep{2002Cappellari}, \href{https://github.com/sibirrer/hierArc}{\textsc{hierArc}} \citep{2020Birrer_TD4}, \href{https://github.com/numpy/numpy}{\textsc{NumPy}} \citep{numpy}, \href{https://github.com/matplotlib/matplotlib}{\textsc{matplotlib}} \citep{Matplotlib} and \href{https://github.com/scipy/scipy}{\textsc{SciPy}} \citep{scipy}.

    HXY thanks Claudia Pulsoni for providing the ETG catalog from the IllustrisTNG simulation. HXY thanks Phil Marshall, William Sheu, Alessandro Sonnenfeld and Veronica Motta for useful discussions and comments. HXY and SB are partially supported by the Department of Physics and Astronomy, Stony Brook University.
    TT acknowledges support by NSF through grants NSF-AST-1906976, NSF-AST-1836016, NSF-AST-2407277, and from the Moore Foundation through grant 8548. DS acknowledges the support of the Fonds de la Recherche Scientifique-FNRS, Belgium, under grant No. 4.4503.1.
\end{acknowledgements}

\bibliographystyle{aa} 
\bibliography{bibliography}

\begin{appendix}
    \section{Normalization factor, $Z$}\label{apdx:norm_factor}

    In this appendix, we discuss the choice of the normalization factor $Z$ in Eq. (\ref{eqn:norm_factor_introduction}). $Z$ is introduced to preserve the total mass of the mass-density profile $\rho(a_v)$, when varying the intrinsic shapes of these profiles. There are two choices of $Z$, (a) $Z = 1$ and (b) $Z = (pq)^{1/3}$, which maps to different effective spherical profiles, i.e., different monopole moments. The total mass within the effective radius, $r_s$, is
    \begin{equation}
        L_k = \int_{a_v=r_s} \mathrm{d}V \rho(a_v),
    \end{equation}
    where $\mathrm{d}V$ is the volume element in a deformed spherical coordinate with $a_v$ as the effective radius. For $Z = 1$, $L_k$ is not conserved with varying the intrinsic axis ratios $p$ and $q$, and thus need to be normalized. For $Z = (pq)^{1/3}$, $L_k$ is conserved at any effective radius $r_s$.

    Without the loss of generality, we numerically test the effect of these two choices of $Z$ on the surface density profile under axial symmetry and MGE form. We express the intrinsic mass-density profile with MGEs as in Eq. (\ref{eqn:density_mge_expression}). The total mass is then
    \begin{equation}
        L_k = \sum^N_{k=1} \nu_{0k} (\sigma_k \sqrt{2\pi})^3 q_k.
    \end{equation}
    We require the total luminosity to be conserved when the intrinsic axis ratio $q_k$ changes to $q_k'$, and thus we can choose (a) $\nu_{0k}' = \nu_{0k}q_k/q_k'$, or (b) $\sigma_k' = \sigma_k (q_k/q_k')^{1/3}$. The former corresponds to a choice of $Z = 1$, and the latter corresponds to the choice of $Z = (pq)^{1/3}$.

\begin{figure*}[ht!]
        \centering
        \includegraphics[width=0.85\linewidth]{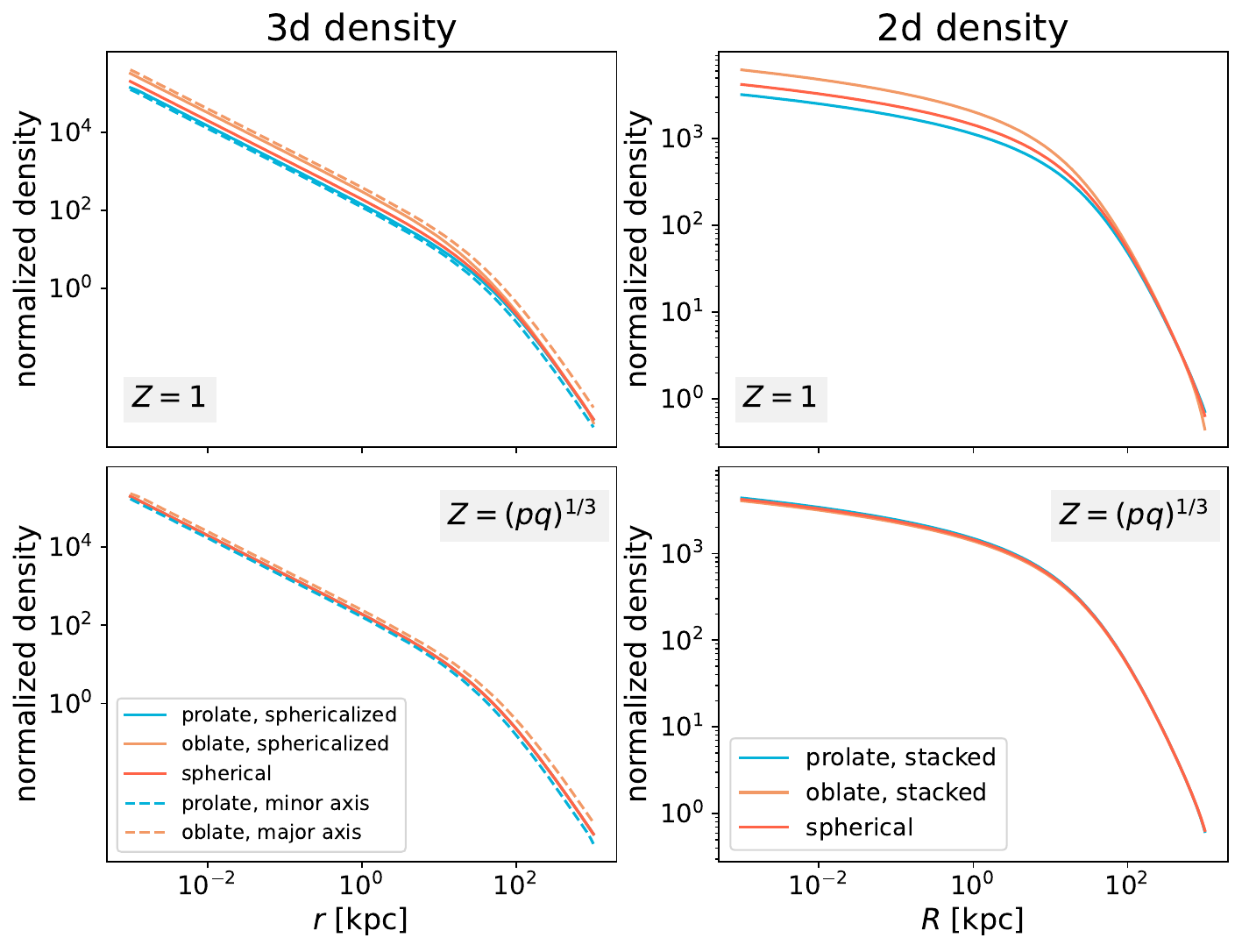}
        \caption{Comparison between the 3D and 2D densities using $Z = 1$ and $Z = (pq)^{1/3}$, under axial symmetry and MGE formalism. Left: 3D density of a NFW profile with different intrinsic axis ratios. The oblate profile has an intrinsic axis ratio of 0.5, and the prolate profile has an intrinsic axis ratio of 1.6. The solid lines are the sphericalized (averaged on a sphere) density profiles. The dashed lines are the 3D density profiles along the symmetry axis of axisymmetric profiles. Right:  Stacked surface density obtained by projecting the 3D densities on the left column 1000 times. }
        \label{fig:sphericalizing_comparison}
    \end{figure*}

    We test whether the surface density profile of an axisymmetric density profile is biased, comparing to a spherical profile with $q_k = 1$. We assume the orientation of the axisymmetric profile is isotropic on a sphere, namely, with the viewing angle of $\cos \theta \in \mathcal{U}[0, 1]$ and the inclination angle $i = \theta$. We project the intrinsic density profile 1000 times, and stack the surface density profiles. We compare the stacked surface density profile with the surface density profile of a intrinsically spherical density profile. Here we use an example NFW halo with a scale radius of 50 kpc and an overall normalization of 1 $\mathrm{M_\odot} / \mathrm{kpc}^3$. We use an intrinsic axis ratio $q_k = 0.5$ for the oblate case and an intrinsic axis ratio $q_k = 1.6$ for the prolate case. The results are shown in Figure \ref{fig:sphericalizing_comparison}. The left column is a 3D NFW density profile with different intrinsic axis ratios. The right column is the stacked surface density profiles. Under the choice of $Z = (pq)^{1/3}$, the stacked surface density profile of an axisymmetric profile is closer to the spherical case. We conclude that choosing $Z = (pq)^{1/3}$ yields more accurate surface density profile, if we assume the stacked surface density profile of a nonspherical profile should approach the spherical surface density of the same mass.

    \section{The fraction of oblate ETG under a simple model}\label{apdx:oblate_fraction}

    \begin{figure}
        \centering
        \includegraphics[width=0.87\linewidth]{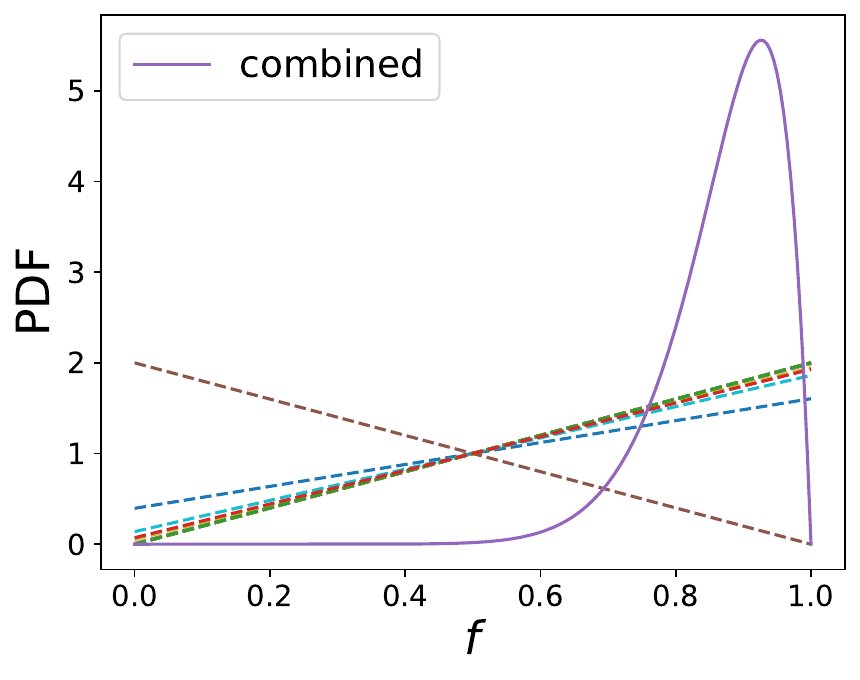}
        \caption{Likelihood function of measured misalignment angles from K24 as a function of the oblate fraction $f$ assuming the model distribution in Eq. (\ref{eqn:apdxB_misalignment_model}).}
        \label{fig:apdxb_oblate_fraction_likelihood}
    \end{figure}

    \begin{figure}
        \centering
        \includegraphics[width=0.84\linewidth]{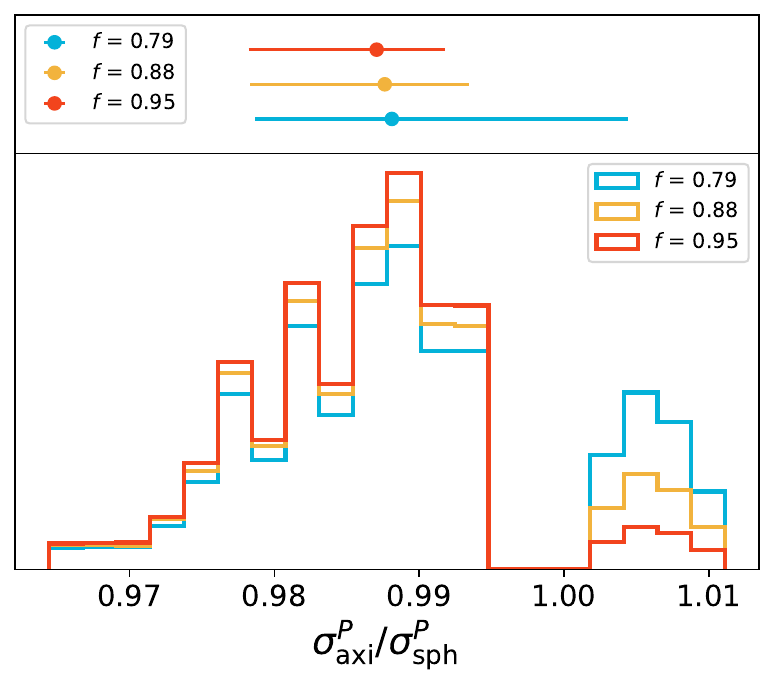}
        \caption{Lower panel:  Normalized distribution of the correction term $\sigma_\mathrm{axi}^P/\sigma_\mathrm{sph}^P$ at $q' = 0.9$ under different values of the oblate fraction $f$. Upper panel:  16th (left end of the errorbar), 50th (the central marker), and  84th (right end of the errorbar) percentile corresponding to the distribution of the correction term. The difference in the median $\sigma_\mathrm{axi}^P/\sigma_\mathrm{sph}^P$ at $f = 0.79$ and $f = 0.95$ is 0.105\%, indicating that the relative change in the corrected velocity dispersion when $f$ changes by two standard deviations is 0.105\%. Therefore, the change in the inferred $\lambda$ and $H_0$ when $f$ changes by one standard deviation is 0.105\%. }
        \label{fig:apdxb_different_oblate_fraction}
    \end{figure}

    Pure oblate galaxies have their photometric major axis and their kinematics major axis fully aligned. Pure prolate galaxies, being very rare in nature, have their photometric major axis and kinematics major axis perpendicular to each other. Galaxies not falling into these two categories are thought to be triaxial. Here we consider a  population composed of oblate and prolate galaxies. The distribution of offsets between the photometric and kinematic axis is
    \begin{equation}\label{eqn:apdxB_misalignment_model}
        P (\Delta \phi\vert f) = f\delta(\Delta\phi) + (1-f)\delta(\Delta\phi - \pi/2),
    \end{equation}
    where $f$ is the fraction of oblate galaxies, and $\Delta\phi$ is the misalignment angle. Taking into account the measurement uncertainty on $\Delta\phi$, the likelihood function is
    \begin{multline}\label{eq:misalignment_likelihood}
        L(\Delta\phi\vert f) = \prod^N_i \frac{1}{\sqrt{2\pi\sigma_{\Delta\phi_i}^2}} \times \\  \left[f\exp(-\frac{{\Delta\phi_i}^2}{2\sigma_{\Delta\phi_i}^2}) + (1-f)\exp(-\frac{{(\Delta\phi_i - \pi/2)}^2}{2\sigma_{\Delta\phi_i}^2}) \right],
    \end{multline}
    where $N$ is the total number of lenses with measured misalignment angles.

    \begin{table}
        \centering
        \begin{tabular}{lcc}
            \hline
            Object         & $\Delta\phi$ ($^\circ$) & $\sigma_{\Delta\phi}$ ($^\circ$)\\
            \hline
            SDSSJ0029-0055 & 31.0                    & 30.0                             \\ SDSSJ0037-0942 & 28.0                    & 4.0                                \\   SDSSJ0330-0020 & 34.0 & 10.0   \\ SDSSJ1112+0826 & 3.0 & 3.0 \\   SDSSJ1204+0358 & 14.0 & 11.0   \\ SDSSJ1250+0523 & 71.0 & 7.0 \\   SDSSJ1306+0600 & 35.0 & 13.0   \\ SDSSJ1402+6321 & 19.0 & 8.0 \\   SDSSJ1531-0105 & 9.0 & 30.0   \\ SDSSJ1538+5817 & 19.0 & 30.0 \\   SDSSJ1621+3931 & 9.0 & 4.0   \\ SDSSJ1627-0053 & 14.0 & 13.0 \\   SDSSJ1630+4520 & 5.0 & 9.0   \\ SDSSJ2303+1422 & 12.0 & 30.0 \\
        \end{tabular}
        \caption{Misalignment angle between the kinematics major axis and the photometric major axis of the lens galaxies from K24.}
        \label{tab:apdxb_misalignment_angle}
    \end{table}

    Feeding in the misalignment angle measured in \citet{2024Knabel} using the Keck Cosmic Web Imager (KCWI) IFU data of a subsample of the SLACS lenses, as listed in Table \ref{tab:apdxb_misalignment_angle}, we model the likelihood as a function of $f$ for individual lenses (dashed lines) and the entire dataset using Eq. (\ref{eq:misalignment_likelihood}), as shown in Figure \ref{fig:apdxb_oblate_fraction_likelihood}. Marginalizing over the combined distribution of $f$ as a weight, we then add the two curves in Figure \ref{fig:bias_axi_vs_sph} to obtain Figure \ref{fig:bias_axi_vs_sph_combined}, from which one obtains a prediction of the correction term in the kinematics modeling.

    We note that the choice of $f$ does not impact significantly the distribution of the correction term $\sigma_\mathrm{axi}^P/\sigma_\mathrm{sph}^P$ under a certain projected axis ratio $q'$. 
    To test this, we model the combined correction term under a randomly chosen projected axis ratio using different single values of $f$ instead of integrating over $f$. Here we choose $\beta=0$, and $q' = 0.9$. 
    For $f$, we choose $f = \{0.79, 0.88, 0.95\}$, corresponding to the 16-th, 50-th and 84-th percentile of our $f$ shown in Figure \ref{fig:apdxb_oblate_fraction_likelihood}. The lower panel of Figure \ref{fig:apdxb_different_oblate_fraction} shows the distribution of the correction term at the chosen  $q'$, while the 16-th, 50-th and 84-th percentiles of the distributions are plotted in the upper panel. 
    We calculate the difference in the median of $\sigma_\mathrm{axi}^P/\sigma_\mathrm{sph}^P$ at $f = 0.79$ and $f = 0.95$. We find $\sigma_\mathrm{axi}^P/\sigma_\mathrm{sph}^P (f = 0.79) - \sigma_\mathrm{axi}^P/\sigma_\mathrm{sph}^P (f = 0.95) = 0.105\%$. 
    We conclude that when the oblate fraction $f$ changes by one standard deviation, the change in the inferred mass-sheet parameter $\lambda$ and the Hubble constant $H_0$ is 0.105\%, which is negligible under the current expected measurement uncertainty of $H_0$.

\end{appendix}

\end{document}